\documentclass{aa}  

\usepackage{graphicx}
\usepackage{txfonts}
\usepackage{comment}
\usepackage{amssymb}
\usepackage{tabularx}
\usepackage[normalem]{ulem}% Extra maths symbols
\usepackage{amsmath}
\usepackage{cases}
\usepackage{lineno}
\usepackage{subfigure}
\usepackage[colorlinks=true, linkcolor=blue, citecolor=blue, urlcolor=blue]{hyperref}
\newcommand{\unit}[1]{\boldsymbol{\rm{\hat{#1}}}}

\newcommand{\obs}{_{\rm obs}}

\begin{document}

   \title{Polarization of impulsive relativistic jets propagating in a stratified medium}

   %\subtitle{I. Overviewing the $\kappa$-mechanism}
%\author{Gal Birenbaum}

   \author{G. Birenbaum
          \inst{1,2}\thanks{Corresponding author: birenbaumgal@gmail.com}
          \and
         P. Beniamini\inst{1,2,3}%\fnmsep\thanks{Just to show the usage
          %of the elements in the author field}
          \and
          J. Granot\inst{1,4,3}
          }

   \institute{Astrophysics Research Center of the Open University (ARCO), The Open University of Israel, P.O. Box 808, Ra’anana 4353701, Israel
              %\email{birenbaumgal@gmail.com}
         \and
             Department of Natural Sciences, The Open University of Israel, P.O. Box 808, Ra’anana 4353701, Israel
             %\email{c.ptolemy@hipparch.uheaven.space}
             %\thanks{The university of heaven temporarily does not
%                     accept e-mails}
        \and
            Department of Physics, The George Washington University 
Washington, DC 20052, USA
\and
Astroparticule et Cosmologie (APC)%\\ 
CNRS – Université Paris Cité, F-75013 Paris, France             }

   %\date{Received September 15, 1996; accepted March 16, 1997 }

% \abstract{}{}{}{}{} 
% 5 {} token are mandatory
 
  \abstract
{Gamma-ray bursts (GRBs) and at least some tidal disruption events (TDEs) are highly energetic transients involving impulsive relativistic jets. %Such impulsive relativistic jets are also thought to be present in some tidal disruption events (TDEs). 
%\pb{These first two sentences sound a little like you are saying ``GRBs are very interesting. TDEs also exist". I would rephrase to something like: ``Gamma-ray bursts (GRBs) and at least some tidal disruption events (TDEs) are highly energetic transients involving impulsive relativistic jets."}
%and as 
As these jets
propagate into their
surrounding media, they drive a relativistic forward shock into it, which radiates multi-wavelength polarized synchrotron emission, known as an afterglow. Analyzing the afterglow light curve along with its polarization curve can shed light on the geometrical properties of the GRB or TDE system such as the magnetic field structure behind the shock, the jet angular structure and even distinguish between proposed scenarios of delayed radio emission seen in some TDEs. %\jgout{One of the progenitor systems of these events are the cores of collapsed stars, which can produce} \jgout{brighter afterglows that dominate the sample of GRB afterglows for which polarization was measured.} \jgout{These jets can propagate into stellar wind, rather than a constant density inter-stellar medium (ISM),} \jgout{which can affect their afterglow signature. However, most afterglow polarization modelsdo not account } \jgout{for this possibility.} 
While most afterglow polarization measurements are for long GRBs, whose massive star progenitor winds are expected to form a stratified external medium, the polarization from such environments is largely unexplored. %\sout{Such environments have also been shown to}
Similarly, stratified media surround jetted TDEs.
In this work we explore how shock propagation into a stratified medium affects the observed afterglow polarization from impulsive relativistic jets. 
%and 
We find that for both top hat and structured jets, in most cases the polarization levels are reduced and its temporal signature evolves on longer time scales compared to a uniform external medium. %\sout{when models accounting for stratified media}. 
%are applied to both top hat and structured jets. 
The polarization peaks at times close to a geometrical break in the light curve, indicating measurements during this time probe the maximal levels of polarization possible for the system. In addition, observationally capturing the polarization evolution time scales can assist in constraining our viewing angle and the external medium stratification.
%the viewing angle to the system.
Such composite models, that account for more realistic systems that are motivated by light curve fittings, will allow us to complement afterglow light curves better and promote joint modeling of these observables. }
   \keywords{Polarization --
                Relativistic processes --
                Shock waves --
                Gamma-ray burst: general
               }

   \maketitle
%
%-------------------------------------------------------------------

\section{Introduction}
\label{sec:intro}

Gamma-ray bursts (GRBs) are an extreme astrophysical transient phenomena, that are usually first detected with a brief flash of $\gamma$-rays, that are then followed by a long-lasting multi-wavelength emission phase.
The duration of the prompt $\gamma$-ray phase has a bi-modal distribution \citep{Kouveliotou1993}, which is mostly attributed to the system in which the GRB originates. Short GRBs, classified as bursts with $T_{90}<2\text{ sec}$, have been theorized and confirmed to be primarily connected to binary compact object mergers involving two neutron stars or a neutron star and a black hole \citep{Eichler1988, Narayan1992, Berger2013, Tanvir2013, Abbott2017, Mooley2018, Nakar2020}. The isotropic equivalent energy in $\gamma$-rays for short GRBs usually lies within the range $10^{50}-10^{52}\text{ erg}$. Long GRBs, with prompt emission durations $T_{90}>2\text{ sec}$,  mostly originate from the core-collapse of massive stars \citep{MacFadyen1999}. Such models were confirmed by observationally associating supernovae to long GRBs \citep{Galama1998,Woosley2006}. The corresponding isotropic equivalent energy in $\gamma$-rays for such bursts lies within the range $10^{52}-10^{54}\text{ erg}$, higher than the energy associated with short GRBs \citep{Nakar2007}.

Both types of
systems eventually leave behind a rotating, accreting compact object, which can launch a relativistic jet \citep[e.g.][]{Blandford1977}, where the emitted prompt $\gamma$-ray emission is due to internal processes within the jet. When these jets break out of their confining media (progenitor star or circum-merger medium), they interact with the ambient medium and drive a relativistic shock into it, which accelerates charged particles \citep{Costa1997,Stanek1999, Nousek2006} and generates magnetic fields \citep{Medvedev1999,2005ApJ...618L..75M}. The outcome of this is a multi-wavelength long lasting synchrotron emission phase, known as the afterglow \citep{Paczynski1993, Sari1998}. 

Synchrotron emission is linearly polarized by nature \citep{Rybicki1979} and afterglow polarimetry in GRBs has demonstrated modest degrees of polarization in the optical ($\lesssim 10\%$; e.g., \citealt{Covino2004, Wiersema2014} and the long GRB 180720B presented in \citealt{Arimoto2024}), in the range of percent to subpercent in the radio band (long GRB 171205A; \citealt{Laskar2020ApJ}, long GRB 260310A; \citealt{Christy2026}), and recent observations managed to put an upper limit of $13.8\%$ on the X-ray polarization of the long GRB 221009A, the brightest observed GRB of all times \citep{Negro2023,Burns2023}. While polarization measurements associated with the reverse shock, which can probe the ejecta, can be diverse (e.g. up to $28\%\pm4\%$ at $240\,\text{s}<t<323\,\text{s}$ in the optical from the long GRB 120308A; \citealt{Mundell2013}, vs sub-percent levels, $0.87\pm0.13\%$ to $0.60\pm0.19\%$ from $2.2\,$hr to $5.2\,$hr after the GRB, in the radio at $97.5\,$GHz from the long GRB 190114C; \citealt{Laskar2019ApJ}), polarization measurements associated with the forward shock, probing the shock-generated magnetic field, are mostly $\lesssim10\%$.  Since the polarization level of most of these sources is low, efficiently measuring polarization in GRB afterglows usually requires a high photon count, i.e., a bright source. Long GRBs tend to provide such bright sources and indeed they dominate the afterglow sample for which polarization was sampled (e.g., all events mentioned in the review by \citealt{Covino2016} are long GRBs, apart for GRB 000301C, classified as short/intermediate duration, see \citealt{Jensen2001}). 

In such long bursts, which arise from the core-collapse of a Wolf-Rayet star \citep{MacFadyen1999}, 
the afterglow shock propagates into the stellar wind, with a power-law density profile behaving as $\rho\propto r^{-k}$ (with $k=2$ for a steady wind), which is ejected during the final evolutionary stages of these stars \citep{Chevalier1999,Chevalier2000,Ramirez-Ruiz2001}.
The interpretation of the measured polarization in unison with the afterglow light curve can be indicative of the jet structure \citep{Birenbaum2024,Birenbaum2026}, which reflects the processes the jet has undergone before breaking out of its confining medium \citep{Beniamini2020, Gottlieb2021, Beniamini2022}. At the same time, most standard polarization and light curve afterglow modeling tools only account for a uniform density medium (e.g., \citealt{Rossi2004, Ryan2020, Birenbaum2021}), which are appropriate for describing the forward shock plunging into the inter-stellar medium (ISM), and reflect the expected conditions for short GRBs \citep{Nakar2007}. However, these may be inappropriate for many long GRB cases (e.g., \citealt{Schulze2011}).

The GRB afterglow formalism has been found to also effectively describe the long-lived radio emission observed in some tidal disruption events (TDEs). This emission is associated with powerful outflows, and there is growing evidence that, at least in some cases, these outflows come in the form of relativistic jets (e.g., \citealt{2020NewAR..8901538D,Beniamini2023, Sfaradi2024,Cendes2026}).  At the point of writing, modest levels of optical polarization have been measured from two jetted TDEs \citep{Wiersema2012,Wiersema2020}. Such measurements can probe the emission mechanism and jet geometry when properly interpreted. As the environments around these systems vary, model flexibility in the density profile the jet probes is required to effectively explain the observations. One such example includes the inference of an off-axis relativistic jet for the TDE AT 2018hyz, whose radio observations are best fitted with a power-law external medium density profile with power-law index close to $1$ \citep{Sfaradi2024}. Another unique example is that of the longest GRB recorded up to the time of writing, GRB 250702B \citep{Neights2025b,Neights2025a,Levan2025,Neights2026}, that presents a detailed multi-wavelength afterglow (e.g., \citealt{Levan2025,Carney2025,OConnor2025}). An interesting interpretation of this event is that of an intermediate-mass black hole (IMBH) disrupting a star, also known as mTDE, resulting in electromagnetic phenomena associated with a relativistic jet \citep{Granot2026}. In \citet{Granot2026}, the authors discuss that when such a black hole is offset from the galactic center, it is expected to accrete the ISM in a quasi-spherical Bondi-like flow, inducing an external medium with a power-law density profile behaving as $\rho\propto r^{-3/2}$. When applying afterglow-like models to the multi-wavelength observations of the event, the authors fit a value of $k=1.67\pm 0.17$, in agreement with the above-mentioned theoretical expectation for such events.

The degree to which a stratified medium affects the observed polarization in impulsive jet afterglows was considered by \citet{Lazzati2004} for on-axis top hat GRB jets, where a slower evolution of the polarization signature is noted. This was later expanded to off-axis top hat jets in \citet{Lan2023}, where the impact of including the possibility of a stratified medium and the equal arrival time surface (EATS) on the observables is explored. However, the top hat jet scenario does not encapsulate the fact that impulsive jets can develop extended angular structures as they drill their way out of their confining medium. In addition, previous works do not take into account the observational indications that afterglow shock-generated magnetic fields posses structures close to 3D isotropic \citep{Corsi2018,Stringer2020,Gill2020,Teboul2021,Arimoto2024}. 

In this paper we use the formalism presented in \citet{GG18} and \citet{Birenbaum2024,Birenbaum2026} to demonstrate how the observed polarization signature changes when the power-law index $k$ of the external density profile the forward shock enters into varies. 
The extension of the model to non-uniform media will allow us to better describe the class of GRBs that are best sampled in the polarized regime due to their brightness, as well as accounting for the environments TDEs occur in. We generalize classical afterglow models for top hat jets propagating in uniform density media (with $k=0$) to general power-law external media density profiles and explore the impact on 
structured jets with a core and power-law wings, viewed off-axis.
Finally, we analyze the trends seen in the light and polarization curves and explore the relation between them to emphasize the importance of joint analysis.

%--------------------------------------------------------------------
\section{Methods}
\label{sec:Methods}

\begin{table}[h!]
 \caption{Afterglow model parameters considered in this work.}\label{table:parm_space}
\centering
\renewcommand{\arraystretch}{1.4} 
\begin{tabular}{l c c }
\hline\hline
 Parameter & Value \\
 \hline
 $\theta_c$    &$3^{\circ}$ \\
 $\Gamma_c$  & 350\\
 $a=-\frac{d\log E_{\rm k,iso}}{d\log\theta}|_{\theta>\theta_c}$ 

 &  Top hat, 1, 3.5  \\
$b=-\frac{d\log(\Gamma_0-1)}{d\log\theta}|_{\theta>\theta_c}$ 

&0  \\
 $E_{\text{c}}$ [erg]& $10^{54}$  \\
 $n_{0}$ [cm$^{-3}$]&  1 \\
 $k=-\frac{d\log n(R)}{d\log R}$
 
 & 0, 1, 2   
 \\
 $R_0$ [cm] & $10^{18}$ \\
 $\nu\obs$ [Hz] & $10^{15}$ (PLS G) \\
  $p$ &  $2.5$ \\
 $\epsilon_\text{e}$   & $10^{-2}$  \\
 $\epsilon_\text{B}$&  $10^{-4}$ \\
  $\chi_\text{e}$&  1 \\
$d_{L}$ [cm] 
& $10^{28}$ \\
z 
& 0.54 \\

\hline
\end{tabular}
%\tablefoot{
% The produced R-band light and polarization curves are presented in Fig. \ref{fig:AT2021lfaXi0AndXi075Lim}. The dirty fireball model is adapted from \citet{Li2024}.
%}
\end{table}

The calculations presented in this work follow the formulas presented in the methods section of \citet[][B24 hereafter]{Birenbaum2024}. The basic model assumed in %\citet{Birenbaum2024} 
B24 and this work features a 2D axisymmetric relativistic forward shock propagating
into a cold ambient medium with a power-law rest-mass number density profile 
\begin{equation}
 n(R)=n_0\left(\frac{R}{R_0}\right)^{-k}\equiv AR^{-k},
 \label{eq:density_profile}
\end{equation}
where $n_0$ signifies the number density the shock encounters at the radius $R_0$ and $k$ is the power-law index of the external density profile. Following \citet{GG18}, the jet dynamics are assumed to be locally spherical (neglecting lateral dynamics) and the emission is calculated from a 2D surface associated with the afterglow shock front. 
The structured jet models explored in this work feature an angular profile that is expressed in terms of the distribution of the isotropic equivalent kinetic energy and initial Lorentz factor, which are described by $E_{\rm k,iso}=E_c\Theta^{-a}$ and $\Gamma_0-1=(\Gamma_c-1)\Theta^{-b}$ where $\Theta=[1+(\theta/\theta_c)^2]^{1/2}$, where $\theta$ is the angle from the jet symmetry axis.
The subscript c denotes properties of the jet core.
The top hat jet model is described by a step function in energy and initial Lorentz factor, where both are uniform up to $\theta_c$ and drop to zero beyond this angle.

The numerical setup used in this work to carry out the calculations is similar to that presented in B24. The shock surface is divided into angular cells around the jet symmetry axis, defined by the polar angle $\theta$ and the corresponding azimuthal angle $\phi$. The emitting region right behind the shock radiates synchrotron emission in a shock-generated magnetic field whose comoving direction $\unit{B}'$ and corresponding magnitude $B'$ in each cell are drawn from a probability distribution set by the magnetic field stretching factor
%\jgcor{anisotropy} parameter 
$\xi$ (for a more detailed explanation, see Appendices A and B of B24). The evaluated observed luminosity at each angular cell $L_{\nu, \text{iso}}$ can be projected onto a spherical projection of the emitting region that shows the contribution to the observed emission at each point on the grid (see blue lines in Fig. \ref{fig:ProjectionSetup} and panel (a) in Figs. \ref{fig:THJq07WithMapsk0}-\ref{fig:THJq07WithMapsk2}).

Following these initial calculations, we proceed to calculate the Stokes parameters $I_{\nu},Q_{\nu},U_{\nu}$ from each cell. Integration over their contributions from all $\theta$ and $\phi$ provides the overall observed linear polarization and flux.
Full details of the calculation are presented in B24.

\section{Results}
\label{sec:Results}
We study the impact of the external medium density profile on the observed afterglow flux and polarization temporal evolution. We focus on power-law profiles with indices in the range $0\leq k \leq 2$. Results are shown in the optical band ($\nu\obs=10^{15} \text{ Hz}$), which, for the chosen underlying parameters, resides in PLS G ($\nu_{\text{m}}<\nu\obs<\nu_{\textbf{c}}$) throughout. Afterglow model parameters used in this work are listed in Table \ref{table:parm_space}. We explore the impact of including the possibility of a stratified medium for on- and off-axis jets in sections \ref{subsec:OnAxisJets} and \ref{subsection:OffAxisJets} respectively and study the effect it has on both top hat and structured jets. While most of the models consider a random magnetic field that is confined to the plane of the shock ($\xi\rightarrow 0$), we also include a corresponding calculation of a more observationally-motivated magnetic field structure behind the shock, with $\xi=0.75$.\;%\footnote{While \citet{Gill2020} derived $0.57\lesssim\xi_f\lesssim0.89$, this is the value just behind the shock, and $\xi$ increases with the distance behind the shock. The relevant value for a 2D emitting surface that is used here would be the radially averaged value, $\xi_{\rm eff}=b^{1/2}$. That work infers $0.66\lesssim b\lesssim1.49$, which corresponds to $0.81\lesssim\xi_{\rm eff}\lesssim1.22$, consistent with the adapted range presented in B24. The results of \citet{Arimoto2024} suggest that $\xi_{\rm eff}>1$, which would altogether imply $1<\xi_{\rm eff}\lesssim1.22$ or $\xi_{\rm eff}\approx1.2$.}

\subsection{Jets viewed on-axis ($0\leq q<1$)}
\label{subsec:OnAxisJets}

\begin{figure}
	\includegraphics[width=\columnwidth]{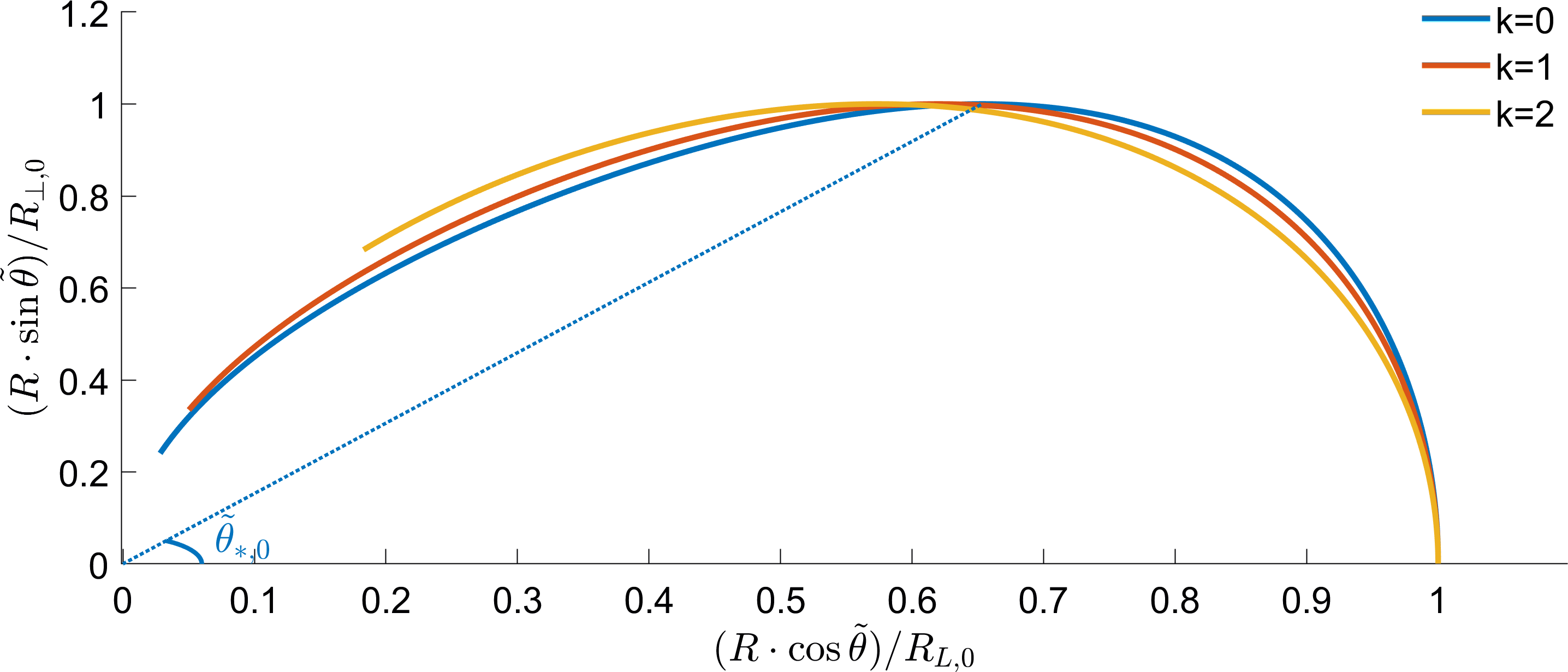}
    \caption{The EATS - the combination of radii and angles that contribute to the observed emission at a given observer time $t\obs$. % for $q=0$. 
    Colors indicate different power-law indices of the external medium density profile $k$. For $k>0$ we re-normalize the shape of the EATS according to the curve shown for $k=0$ in both axes in order to allow for easier comparison.
    }
\label{fig:EATSAllk}
\end{figure}
\begin{figure}
  \centering
 
  {\includegraphics[width=0.9\columnwidth]{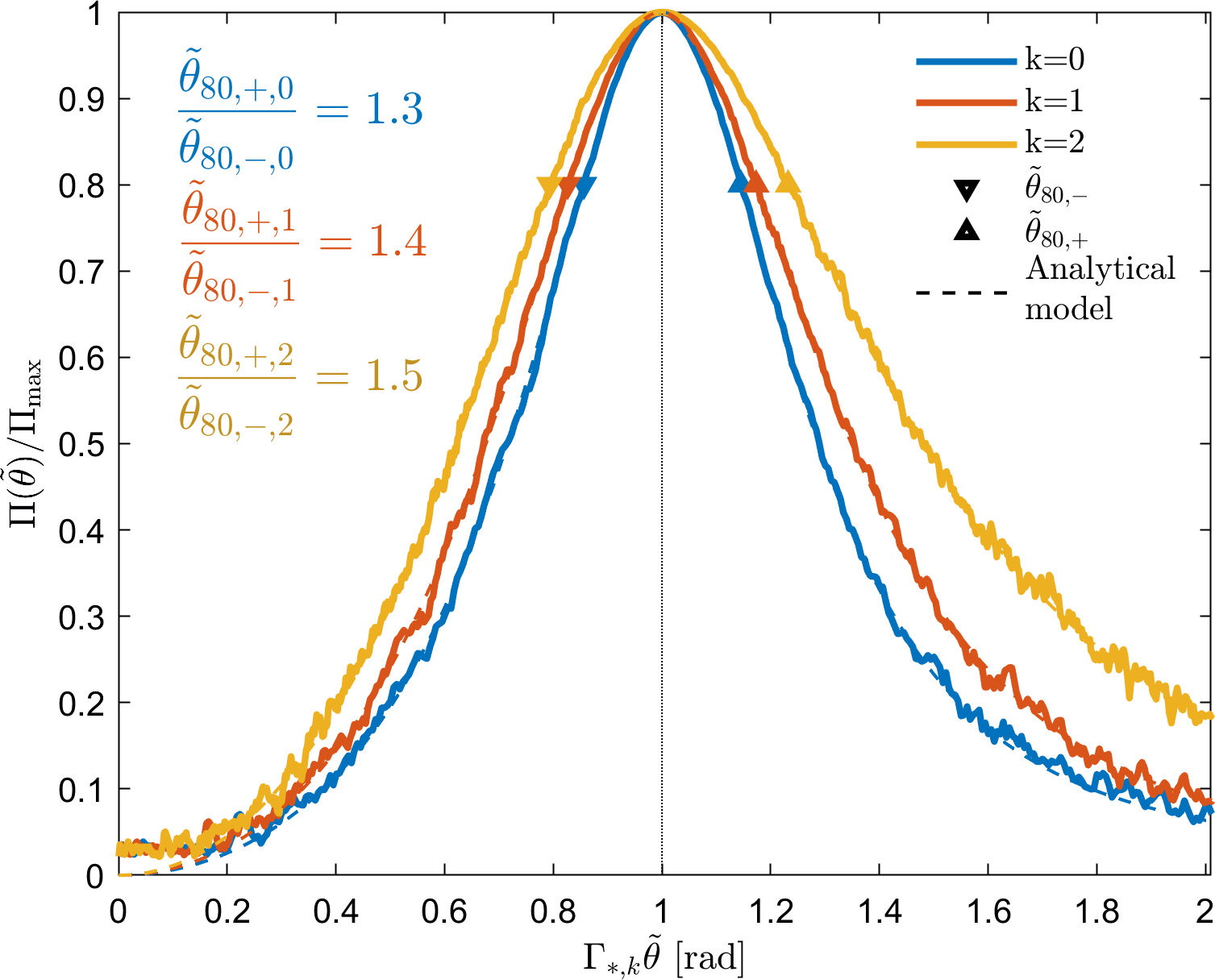} }

  \caption{Angular contribution to the observed polarization for a spherical flow, also applicable for a top hat jet, observed on-axis when $\Gamma\gg\frac{1}{\theta_c-\theta\obs}$ and $\xi\rightarrow 0$, for varying values of $k$, normalized according to $\Pi_{\text{max}}=\frac{p+1}{p+7/3}$ \citep{Rybicki1979}, plotted as function of $\Gamma_{*,k}\tilde{\theta}$. The contribution to the polarization peaks at the peak of the EATS height, where $\Gamma_{*,k}\tilde{\theta}=1$, independently of $k$. In dashed thin lines we plot the corresponding analytical models from \citet{Sari1999Pol} and \citet{Granot2003}.
  }
 
\label{fig:RingAllk}
\end{figure}

\begin{figure*}
  \centering
  \subfigure%(b)
  {\includegraphics[width=0.8\textwidth]{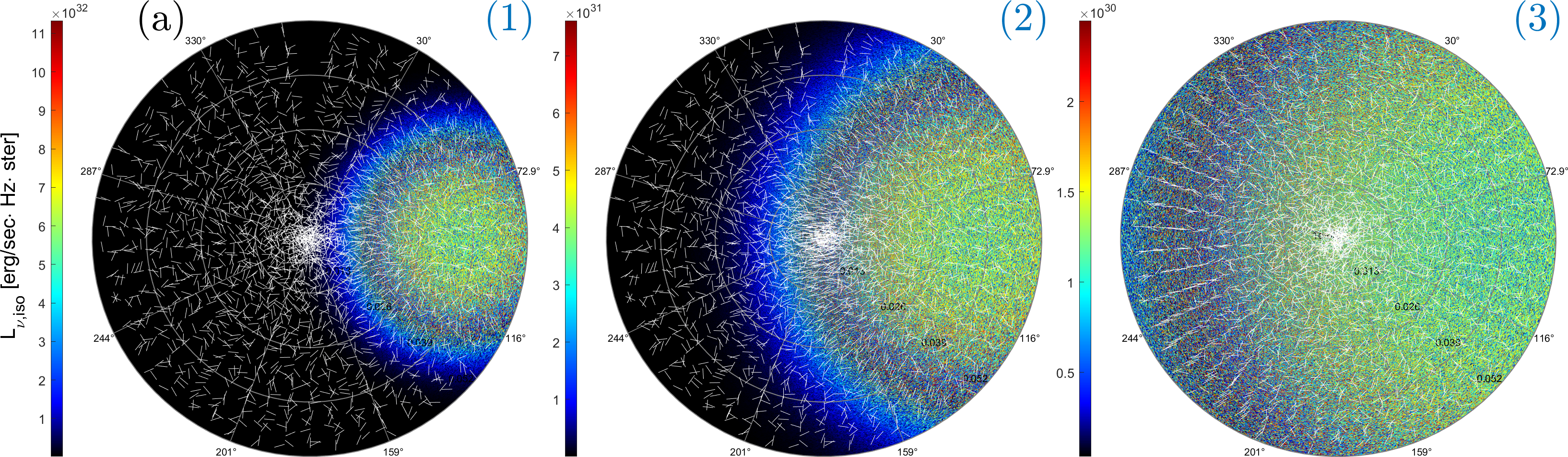}} 
  \\
  \subfigure%(d)
  {\includegraphics[width=0.8\textwidth]{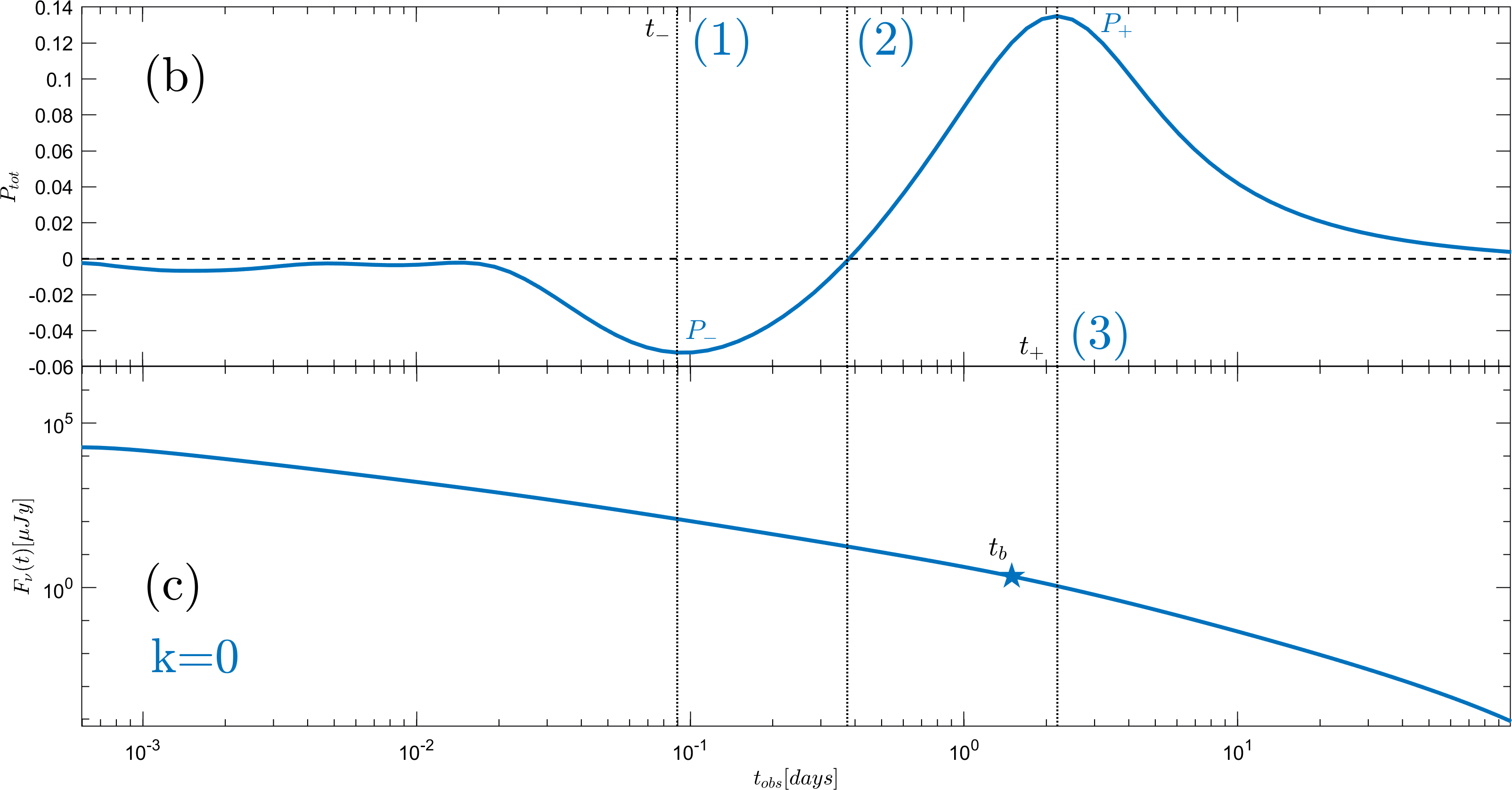}}

  \caption{Polarization signature of a top hat jet, observed at $q=0.7$ with $\nu\obs=10^{15} \text{ [Hz]}$ and a random magnetic field confined to the shock plane ($\xi\rightarrow 0$), for $k=0$. \textit{Panel (a)}: Angular polarization maps of the emitting region (plotted using the spherical projection in Fig.~\ref{fig:ProjectionSetup}, blue lines), where the radius of the map corresponds to the polar angle $\theta$ from the jet symmetry axis that is located at the map center and the edge of the map is at $\theta_c$. The color map represents the contribution to the observed luminosity at each cell ($L_{\nu,\text{iso}}=D^3(\tilde{\mu})L'_{\nu',\text{iso}}$; for additional details see \citealt{GG18} and B24), where it is ring-like around the line of sight (located to the right of the jet axis) and presents a mixture of magnitudes according to the local direction of the magnetic field. White lines represent the direction of the local polarization vector in the observer frame, and for $\xi\rightarrow 0$, it is mostly radially ordered across this bright ring. Frames (1)-(3) depict the system at times when the polarization reaches its negative peak, zero and positive peak respectively. 
  \textit{Panel (b)}: Time evolution of the observed polarization. This is computed by combining the local direction of the polarization vector in each angular cell with the contribution to the flux to produce flux weighted Stokes parameters, which are then integrated over the map to produce the observed polarization level. A change in sign stands for a $90^{\circ}$ rotation of the polarization vector. For additional details, see the Methods section in B24. \textit{Panel (c)}: Light curve of the model presented above. The jet break is marked with a blue star and is close in time to the second polarization peak (frame (3)).}
 
\label{fig:THJq07WithMapsk0}
\end{figure*}

\begin{figure*}
  \centering
  \subfigure%(b)
  {\includegraphics[width=0.8\textwidth]{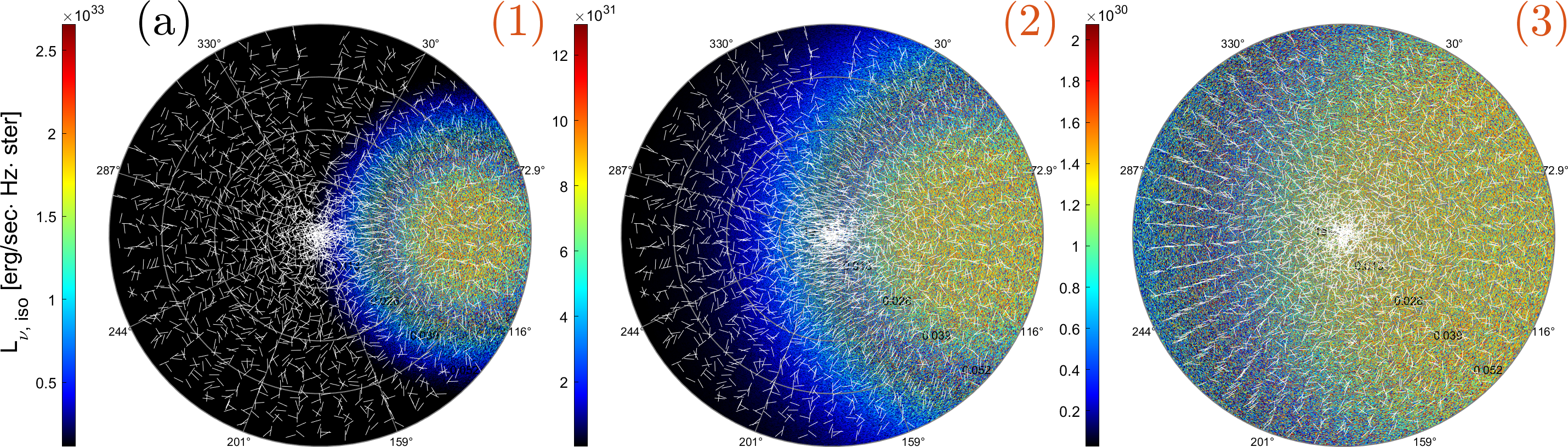}} 
  \\
  \subfigure%(d)
  {\includegraphics[width=0.8\textwidth]{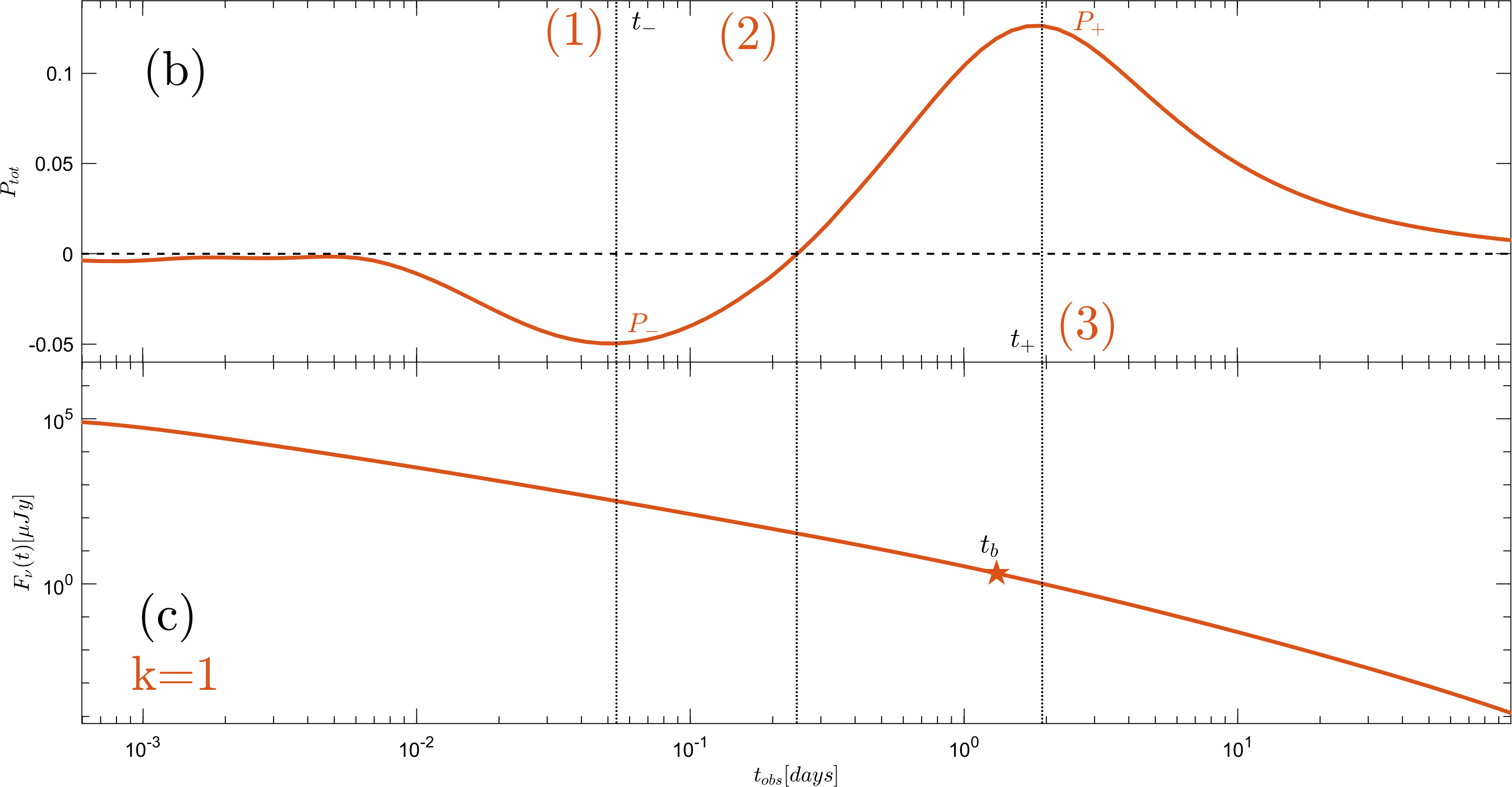}}

  \caption{Same as Fig. \ref{fig:THJq07WithMapsk0} for $k=1$.}
 
\label{fig:THJq07WithMapsk1}
\end{figure*}

\begin{figure*}
  \centering
  \subfigure%(b)
  {\includegraphics[width=0.8\textwidth]{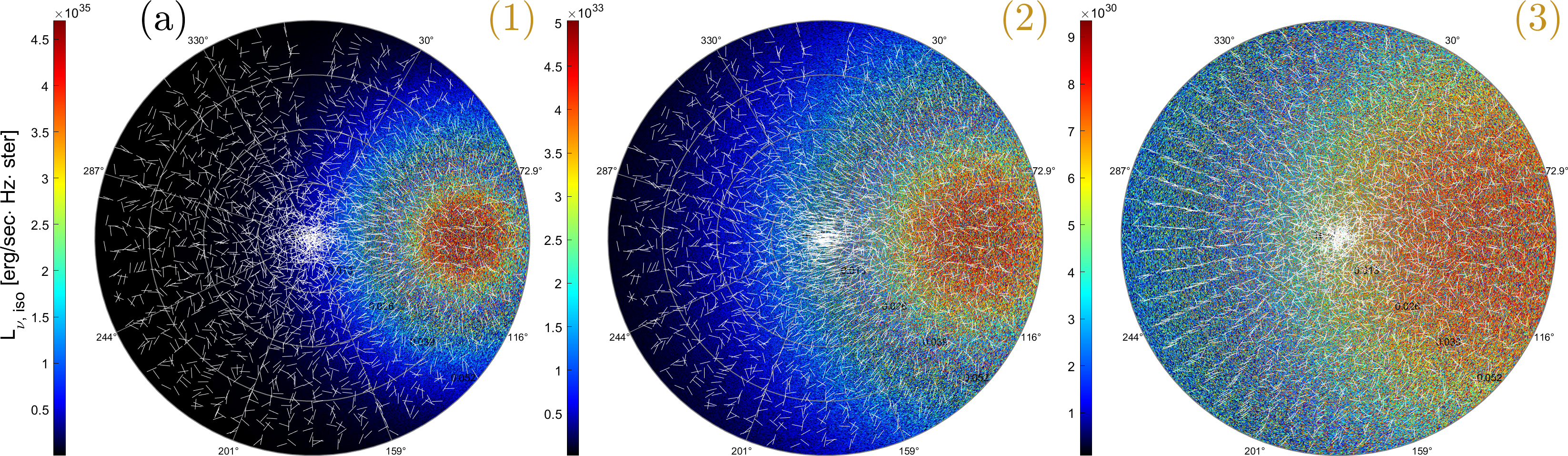}} 
  \\
  \subfigure%(d)
  {\includegraphics[width=0.8\textwidth]{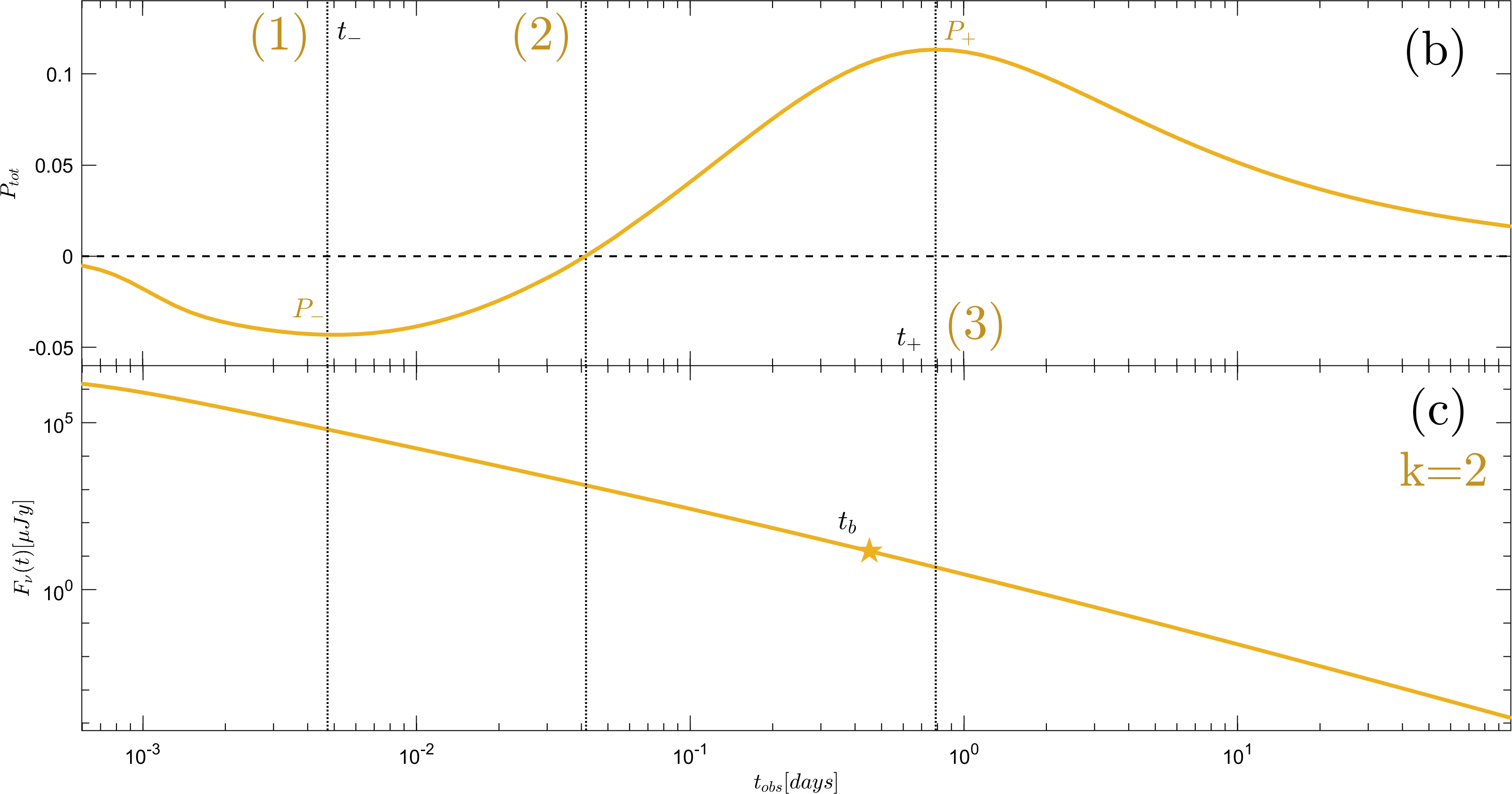}}

  \caption{%\sout{Polarization signature of a top hat jet, observed at $q=0.7$ at $\nu\obs=10^{15} \text{[Hz]}$ with a random magnetic field confined to the shock plane ($\xi\rightarrow 0$), propagating into a wind-like external medium with k=2. \textit{Panel (a)}: Angular polarization maps of the emitting region (see description in the caption of Fig. \ref{fig:THJq07WithMapsk0}). Frames (1)-(3) sample the system at time the polarization reaches its negative peak, zero and positive peak respectively. We can see the polarized ring is wider compared to those shown in Figs. \ref{fig:THJq07WithMapsk0}, \ref{fig:THJq07WithMapsk1}. \textit{Panel (b)}: Time evolution of the observed polarization, where a change in sign stands for a $90^{\circ}$ rotation of the polarization vector. We note the temporal evolution of the polarization signature is "smeared" in time compared to those seen in Figs. \ref{fig:THJq07WithMapsk0}, \ref{fig:THJq07WithMapsk1}. \textit{Panel (c)}: Light curve of the model presented above, where the jet break is marked with a yellow star and is close in time to the second polarization peak (frame (3)).}
  Same as Fig. \ref{fig:THJq07WithMapsk0} for $k=2$.}
 
\label{fig:THJq07WithMapsk2}
\end{figure*}

We start by explaining the impact of a power-law external medium density profile on the emitting region that contributes to the observed polarization from a spherical flow, which also holds for an on-axis top hat jet, when $\Gamma\gg\frac{1}{\theta_c-\theta\obs}$. This setup is a simpler example that would later on help in understanding more complex systems. Previous works have shown that for a uniform external density profile with $k=0$, in the declining region of the synchrotron spectrum, the area that contributes most to the observed emission is in the shape of a ring around the line of sight with an angular radius of $\sim\frac{1}{\Gamma}$, due to the combined effect of contribution from the equal arrival time surface (EATS; see blue lines in Fig. \ref{fig:EATSAllk}) and the fact the spectrum is declining (e.g. \citealt{Sari1999Pol, Granot2008, Birenbaum2021}). In addition, the polarization vector is ordered\footnote{With its direction and sign depending on the sign of $\log\xi$, where it is radial (azimuthal) w.r.t the line of sight for a negative (positive) $\log\xi$. For $\log\xi<0$, see panel (a) of Figs. \ref{fig:THJq07WithMapsk0}-\ref{fig:THJq07WithMapsk2}.} across this bright ring, %\jg{rather than saying ``the polarization vector is ordered" (as this ordering is the same for any ring around the line of sight \gb{this is not the case with the code used for this model... the direction of the polarization vector is calculated at each point as a result of the $\xi$ param and the beaming... you can see that in the polarization maps below}) it is more accurate to say that the local polarization is high, as demonstrated in Fig.~\ref{fig:RingAllk}, which corresponds to $\theta'\approx90^\circ$, or the local emission direction being close to the plane of the shock, which is thus viewed close to edge-on making the field projected onto the plane of the sky nearly ordered, leading to $\Pi\approx\Pi_{\rm max}$}
making it the area that can contribute to the observed polarization (e.g. \citealt{Sari1999Pol,Granot2003a, Nava2015a, Birenbaum2021,Shimoda2020}). This is demonstrated in Fig.~\ref{fig:RingAllk}, that plots the flux-weighted polarized fraction, where the addition of the Stokes parameters is done by summing over cells of different $\tilde{\phi}$ for a fixed $\tilde{\theta}$, where $\tilde{\theta}$ and $\tilde{\phi}$ are the polar and azimuthal angles relative to the line of sight, and using $\hat{\tilde{\phi}}$ as the local reference direction for the Stokes $Q_\nu$ and $U_\nu$ parameters.
%relative to the local reference direction, which is radial for the value $\xi\rightarrow 0$ considered here. 
%\jg{This is not clear enough in my opinion, and a little explanation along the following lines could help clarify this to the readers. According to the analytic model for $\xi\to0$ that was presumably used here (this should be stated in the figure caption), the polarization along the image is always in a redial direction relative to its center. Therefore, the addition of the Stokes $Q$ and $U$ parameters should be relative to this local (location-dependent) reference direction, while if added relative to a single global reference direction they would add up to zero. This is the key to the correspondence with the analytic models, and why the different random realizations for the same $\theta$ and different $\phi$ corresponds to an MC version of the analytic local field direction probability  distribution in the analytic models.} 
This quantity is plotted as function of $\Gamma_{*,k}\tilde{\theta}$, where $\Gamma_{*,k}$ is the Lorentz factor at the angle $\tilde{\theta}_{*,k}$ at which the height of the EATS peaks (see Fig. \ref{fig:ProjectionSetup} where these quantities are marked). Plotting the polarized fraction as function of this quantity reflects the state of the polarized region at all observer times where $\Gamma_{*,k}\gg 1$. We note that the angular contribution to the observed polarization peaks at the angle $\tilde{\theta}_{*,k}=\frac{1}{\Gamma_{*,k}}$ where the maximal height of the EATS is reached, and is ring-like around the line of sight at $\tilde{\theta}=0$.%\jg{since for $\Gamma\gg1$, we have $\cos\theta'\approx\frac{1-y}{1+y}$ where $y=(\Gamma\theta)^2$, if the $x$-axis in Fig.~\ref{fig:RingAllk} will be changed to $\Gamma\theta$ then all three corves would coincide. Therefore, the different width of the rings in the current figure reflects the different variation of $\Gamma(\theta)$ along the EATS.}

This ring-like structure of the observed region persists also when the value of $k$ is increased to include stratified media. In Fig. \ref{fig:EATSAllk}, we plot the EATS for values of $k>0$, normalized according to the $k=0$ curve and plotted at the same observer time in order to compare their shapes. While the shapes of all curves are similar, we can see that the height of the EATS peaks at larger values of $\tilde{\theta}$.\,\footnote{which will correspond to larger values of $\tilde{\theta}_{*,k}$ for the non-normalized versions of the EATS.}
A similar effect can also be seen in the analytical expressions for the surface brightness contribution to the afterglow image for different values of $k$ (\citealt{Granot2008}; also see gray lines in Fig. \ref{fig:ProjectionSetup}) and in the analytical solution for the EATS for different values of $k$, presented in Appendix \ref{app:AnalyticalEATS}.
This indicates the width of the bright ring increases with the value of $k$. This is further quantified in Fig. \ref{fig:RingAllk}, where we expand the analysis to $k>0$. The angular contribution to the observed polarization still peaks at $\Gamma_{*,k}\tilde{\theta}=1$. The result of our semi-analytical model is consistent with the analytical models of \citet{Sari1999Pol} and \citet{Granot2003} for a random magnetic field confined to the plane of the shock. %\jg{The explicity equation for a general $\epsilon=\frac{p+1}{2}$ in PLS G is given in Eq.~3 of \citet{Granot2003} and shown for several different values in Fig.~2 of that work, versus $\mu'=\cos\theta'=\frac{1-y}{1+y}$ (i.e. $\left(\frac{1-y}{1+y}\right)^2=\mu^{\prime\,2}$ and $\frac{4y}{(1+y)^2}=1-\mu^{\prime\,2}=\sin^2\theta'$) where $y=(\Gamma\theta)^2$. It is a 1D integral over $\phi$ that is easy to make for a proper comparison} 
We quantify the angular width of the polarized ring by marking for each model the angles at which the polarization reaches $80\%$ of its maximal height with $\tilde{\theta}_{80,\pm,k}$, namely $\Pi(\tilde{\theta}_{80,\pm,k})=0.8\Pi_{\text{max}}$. The fraction of the observed flux within this $\tilde{\theta}_{80,\pm,k}$ ring is $26.1\%$ for $k=0$, $27.7\%$ for $k=1$ and $28.3\%$ for $k=2$. %\jg{It will help to also quote the fraction of the polarized flux coming from the ring, i.e. the flux weighted by the local $\Pi(\theta)$, which is the most relevant for the polarization.}
%\gb{Now this is a bit of a philosophical question: Since this is a system with q=0, integrated polarization is zero hence it'll be zero for all k. If we want to ask how much of the flux in this region is polarized, we can also do that and then there's the question of how much of the flux in this ring is polarized (67\% for k=0), or how much of the total polarization is contained in this ring, i.e. $P_{80}/F_{tot}$ which is 17.5\% for k=0 (whereas $P_{tot}/F_{tot}=37\%$. I would go with zero as we are looking at integrated quantities.}\jg{It in not clear exactly what you mean by $P_{80}$. What I meant by polarized flux is $\int dF_\nu\Pi(\theta)$, and the most relevant thing to quantify is what fraction of this comes from the ring, i.e. $\int_{\theta_{80-}}^{\theta_{80+}} dF_\nu\Pi(\theta)\;/\int_0^\pi dF_\nu\Pi(\theta)$ (note that here $\theta$ is the angle from the line of sight, which is why there is no dependence of the local polarization on the corresponding azimuthal angle $\phi$). If this is indeed a large fraction, then obscuring a significant part of this ring (e.g. by being outside of the edge of a top-hat jet) could induce significant net polarization. This is why it will help to add ti Figs. 3-5 similar color-maps showing the polarized flux in the color-map, i.e. $\Pi(\theta)\times(dF_\nu/dA)$}
  The ratio between these angles (presented in Fig. \ref{fig:RingAllk}) increases with $k$, indicating that the angular width of the polarized bright ring widens with $k$.%\sout{changes with the value of $k$, where the ring is narrower for $k=0$ than it is for $k=2$. However, when the system is observed perfectly on-axis, the overall polarization will be zero, as the setup is completely symmetrical around the line of sight.} 

\begin{figure*}
  \centering
  \subfigure%(a)
  {\includegraphics[width=0.9\columnwidth]{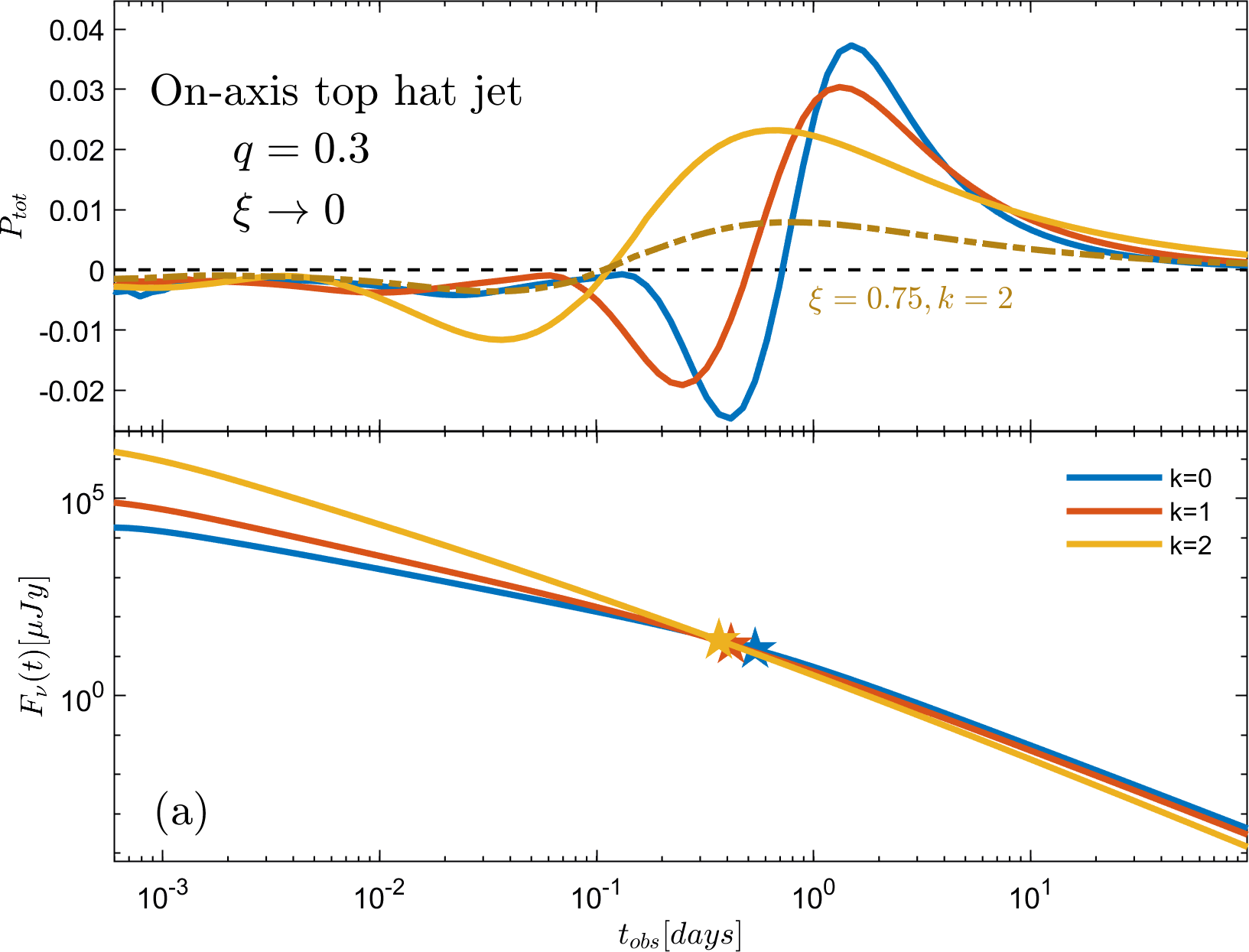} }
  \subfigure%(b)
  {\includegraphics[width=0.9\columnwidth]{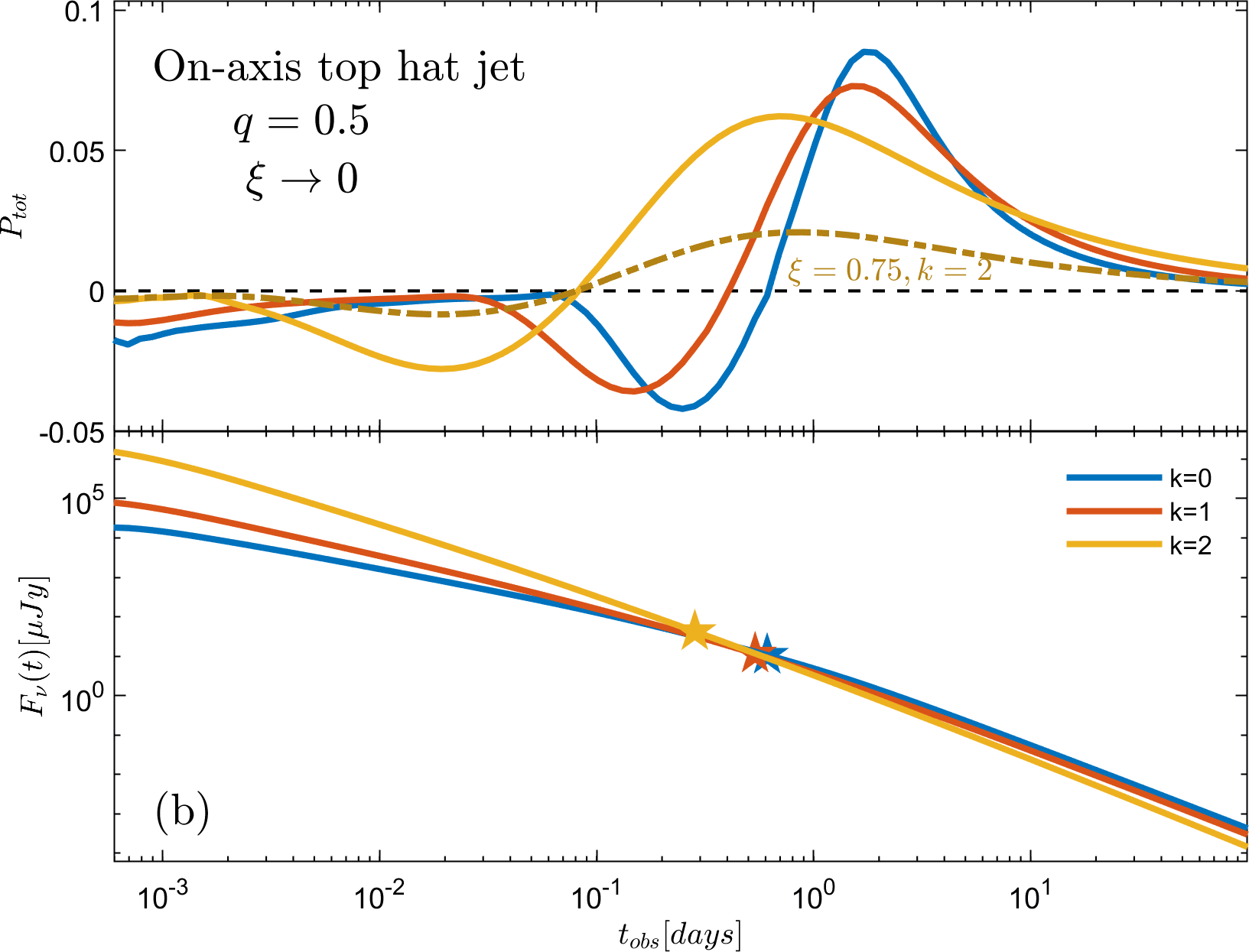}} 
  %\subfigure%(d)
  %{\includegraphics[width=0.9\columnwidth]{PDFTopHatOnAxisTogetherWithXiWithJetBreakWithSubPlot.png}} 
  %\subfigure%(d)
  %{\includegraphics[width=0.9\columnwidth]{PDFTopHatJetq2AllkTogetherWithXiWithJetBreakWithSubPlot.png}} 
  \caption{Observed polarization degree (\textit{upper panels}) and flux (\textit{lower panels}) as a function of time for an on-axis top hat jet model observed at $\nu=10^{15}$ [Hz] (PLS G) with a random magnetic field structure, confined to the shock plane ($\xi\rightarrow 0$). $k$ denotes the power-law index of the external medium density profile. The light curve break is marked with a star in the lower panels. The dash-dotted dark yellow lines present the polarization curves for cases considering a more realistic magnetic field configuration behind the shock with $\xi=0.75$. At the earliest times only a relatively small number of angular cells contribute to the observed emission, leading a small but finite degree of polarization.}
 
\label{fig:THJModelsOnAxis}
\end{figure*}

\begin{figure}

  {\includegraphics[width=0.9\columnwidth]{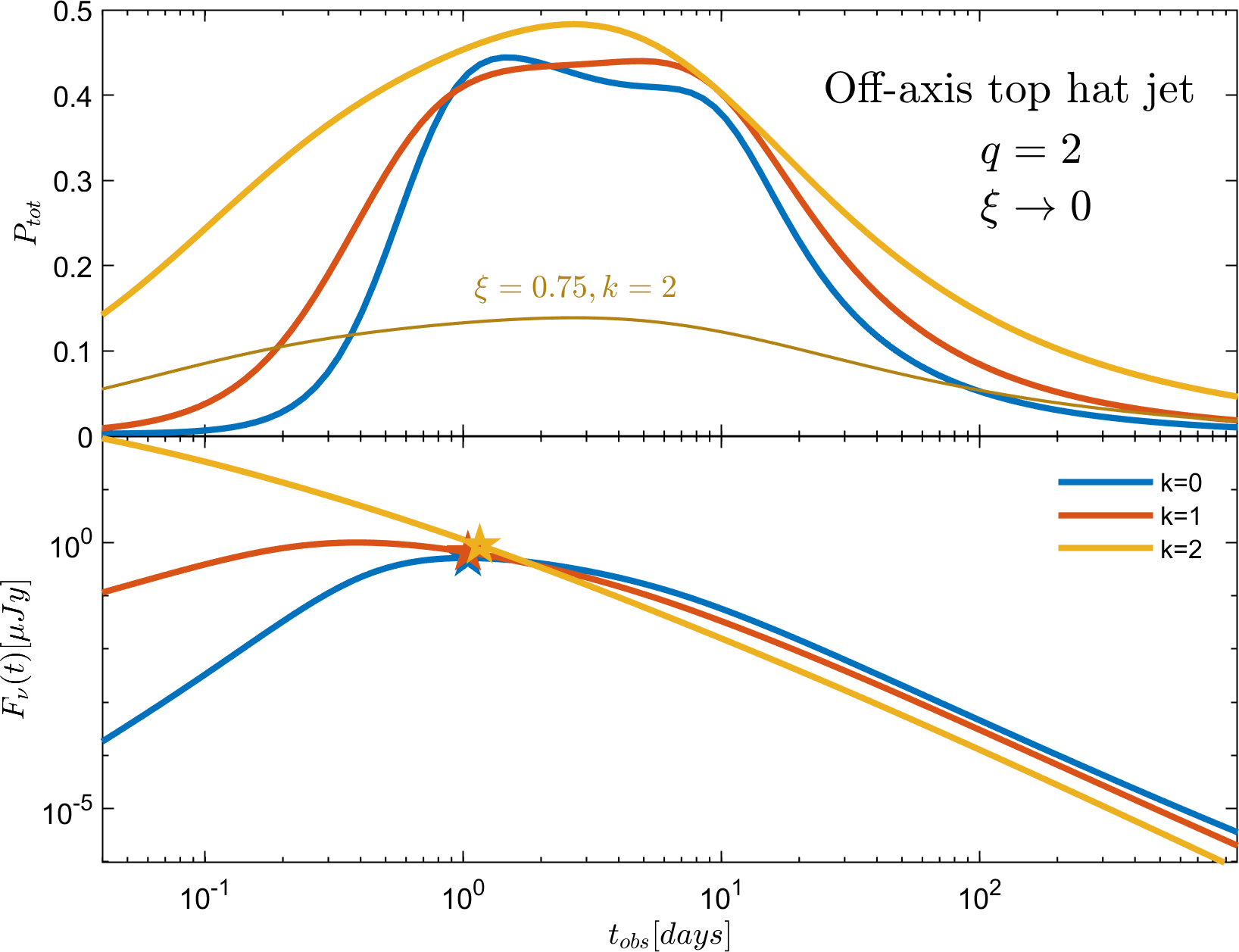}} 
  \caption{Same as Fig. \ref{fig:THJModelsOnAxis} for $q=2$. }
 
\label{fig:THJModelsOffAxis}
\end{figure}

Due to azimuthal symmetry, when the jet system is observed perfectly on-axis, the overall polarization will be zero for a shock-generated field\footnote{for cases considering uniform fields, non-zero observed polarization will be produced even for $q=0$ (e.g., \citealt{Granot2003a, Nava2015a}).}. A misaligned viewing angle introduces asymmetry and hence non-zero polarization. % \gb{The results above hold also for misaligned jets where $\theta$ in Fig. \ref{fig:RingAllk} is replaced with $\tilde{\theta}$.}
%\gb{In order to produce observed polarization in the perfectly on-axis top hat jet scenario, we need to introduce asymmetry to the system in terms of misalignment of the line of sight with the jet symmetry axis.}
We describe next the polarization signature of on-axis top hat jets with $0<q<1$ and $k=0,1,2$. Such polarization signatures has been explored in detail in past works \citep{Sari1999Pol,Ghisellini1999, Granot2003a,Rossi2004,Shimoda2020,Birenbaum2021} and includes two polarization peaks with opposing signs, corresponding to the polarization angle being rotated by $90^{\circ}$ when the polarization degree crosses zero. This polarization signature for a uniform density external medium with $k=0$ and $q=0.7$ is presented in Fig. \ref{fig:THJq07WithMapsk0}\footnote{This plot is similar to Fig. 6 in \citet{Birenbaum2021}}. %\sout{, where panel (a) presents the polarization maps at selected critical times, where the color map shows the contribution to the luminosity $L_{\nu}$ at each point on the angular grid and the white lines represent the direction of the local polarization vector in the observer frame. Panels (b) and (c) show the polarization and light curves respectively, with numbered vertical lines at the critical times.}
%\pb{We don't need to describe the details of the figure panels etc. in the main text - this is done in the caption. It interrupts the flow here. Only mention results / conclusions / insights here.} 
As the blast wave decelerates, the angular size of the bright polarized ring grows and eventually starts disappearing beyond the sharp edge of the top hat jet. %When the line of sight is completely aligned with the symmetry axis of the jet ($q\equiv\frac{\theta\obs}{\theta_c}=0$), the system is entirely symmetrical to the observer, and will produce zero observed net polarization (see illustration in the lower panel of Fig. \ref{fig:RingAllk}). 
When eventually a quarter of the ring disappears beyond the jet edge, the observed polarization experiences its first peak (frame (1) of panels (a) and (b)). As the shock wave decelerates further, it reaches a point in time where only half of the polarized ring is visible, and the contributions from the visible part cancel each other out completely, leading to the observed polarization to be zero (frame (2) of panels (a) and (b)). Following this point, the observed emission and polarization become more dominated by a smaller fraction of the polarized ring. The emerging observed polarization vector is now rotated by $90^{\circ}$, corresponding to a change in the polarization sign. Eventually, only approximately a quarter of the polarized ring is visible, across which the local polarization vector is ordered (frame (3) in panels (a)-(b) of Fig. \ref{fig:THJq07WithMapsk0}). This leads to minimal cancellations and a second polarization peak, which is higher than the previous one due to the emission now being dominated by parts of the polarized ring that are polarized in approximately the same direction \citep{Sari1999Pol,Birenbaum2021}.

The impact %\sout{the power-law external density profile has}
of $k$ on the polarization signature of an on-axis top-hat jet is explored in Figs. \ref{fig:THJq07WithMapsk1}-\ref{fig:THJModelsOnAxis}. %Fig.~\ref{fig:THJModels}, panels (a)-(c). 
The polarization signature of an on-axis, top hat jet, composed of two polarization peaks at opposing signs %\sout{still remains even when the afterglow forward shock propagates into a power-law external medium (see Figs. \ref{fig:THJq07WithMapsk1}-\ref{fig:THJq07WithMapsk2}), consistent with the results shown in \citet{Lazzati2004} and \citet{Lan2023}}
is unchanged for $k>0$ (see also \citealt{Lazzati2004,Lan2023}). However, there are differences in both the overall degree of polarization and the timescales over which it significantly varies. 

The peak degree of 
%the 
polarization 
%peaks 
%\sout{becomes lower when the forward shock propagates into an external medium with a non-trivial power-law profile (with $0<k\leq 2$)}
decreases with increasing $k$. This is because larger $k$ means the polarized ring is wider and takes longer to disappear past the jet edge while inner-more parts still contribute, causing more cancellations in a more symmetrical emitting region.
%\sout{. As stated above, polarization appears in on-axis top hat jets when parts of the polarized ring disappears beyond the the jet edge. When the polarized ring is thinner, parts of it will disappear entirely beyond the jet edge while others remain fully visible, making the observed region highly asymmetrical during the first polarization peak, leading to a higher peak polarization. %\jg{This and the following sentence are hard to follow if this is not demonstrated in a figure}
%However, when the ring is wider, it is not only a single pole of it that starts disappearing beyond the jet edge as the shock decelerates, but additional parts of the ring disappear as well, while inner-more parts still contribute, causing more cancellations in the now more symmetrical emitting region and lower observed polarization levels. This is shown in the polarization maps presented in panel (a) of Figs. \ref{fig:THJq07WithMapsk1}-\ref{fig:THJq07WithMapsk2}.} %\jg{This whole explanation will be greatly improved with a new figure along the lines we discussed. Maybe it would be better to show $\frac{dF_\nu}{d\theta}=2\pi\sin\theta\frac{dF_\nu}{d\Omega}$ rather than $\frac{dF_\nu}{d\Omega}$, and define $\Delta\theta_{50}(t_{\rm obs})$ as the smallest (central) range (between $\theta_{50-}$ and $\theta_{50+}$) contributing $50\%$ of the total flux (or polarized flux); $\Gamma_*\Delta\theta_{50}$ should be constant in the spherical self-similar phase and its value can reflect the effective angular width that may be associated with a ``polarized ring''. For the timescales, } 
This is especially true during the second polarization peak, that is associated with the jet break in the afterglow light curve. %\sout{, marked with a star}\footnote{The association between the the jet break and polarization peak is discussed in \citealt{Sari1999Pol,Granot2003,Birenbaum2021}; B24; \citealt{Birenbaum2026}; we also demonstrate this point in Appendix~\ref{app:TimeRatios}}. %\jg{This does not appear to be clearly demonstrated there - am I missing something?} 
%\sout{While indeed a single lobe of the polarized ring dominates the emission in the wind-like external medium case, since the ring itself is wider, the inner parts of the other ring lobes still remain visible, allowing again for cancellation to take place and reducing the observed polarization during the second polarization peak as well (see frame (3) in panel (a) of Figs \ref{fig:THJq07WithMapsk1}-\ref{fig:THJq07WithMapsk2}).} %\jg{This explanation is not very clear to me, lacking a clear enough figure to support the description in the text. It completely doesn't mention the important dependence of the local polarization on $\theta'$ - the comoving angle to the observer relative to the shock normal; see e.g. Fig.~2 of \citet{Granot2003}} 
%\sout{We find that polarization signatures from impulsive jet afterglow shocks that propagate into power-law media with $0<k<3$\pb{You mention $k=3$, but we only showed $k=0,1,2$.} feature polarization curves with lower magnitude compared to those propagating into a uniform density medium with $k=0$, also in agreement with \citet{Lazzati2004} and \citet{Lan2023}.}
%\pb{The preceding paragraph is a repetition of what we said above - it can be deleted.}
For a simpler comparison between the different models, we also plot light and polarization curves for all values of $k$ considered in this work for values of $q=0.3,0.5$ in Fig. \ref{fig:THJModelsOnAxis}, which show similar trends.

%\gb{you mix here the fact that the time scales are smeared which is robust with the features becoming early or late...}
The time scales over which the polarization signature evolves depend on the stratification of the medium. Figs. \ref{fig:THJq07WithMapsk0}-\ref{fig:THJModelsOnAxis} show a ``smearing" of the polarization signature across longer time scales for higher values of $k$, as the interaction of the jet edge with the bright polarized ring lasts longer due to its increased width.
%Moreover, the large densities the shock goes through at early observer times, cause its early deceleration to lower Lorentz factors (see Fig. \ref{fig:Gamma(t_obs)}).\jg{It seems to me that the dominant effect is the scaling of $\Gamma\propto t_{\rm obs}^{(k-3)/(8-2k)}$ that causes slower evolution for larger $k$. You can try to plot the same polarization curves as a function of $\Gamma_*=(5-5)^{\frac{3-k}{2(4-k)}}\Gamma_L$ instead of $t_{\rm obs}$ and compare the differences there, which may be attributed to the shape of the EATS or ring.} 
In addition, the scaling of the Lorentz factor with time $\Gamma\propto t_{\rm obs}^{(k-3)/(8-2k)}$ shows slower evolution of $\Gamma$ for larger $k$. %\sout{The lower the Lorentz factor is, the longer it takes the relativistic shock to further decelerate, thus contributing to the ``smearing" effect of the polarization signature across longer time scales for values of $k>0$.}
Since the polarization evolution depends directly on $\Gamma$, determining the visible region of the jet at each point in time, shallower $\Gamma$ evolution corresponds to ``smeared" temporal evolution.

While the results above are shown for a random magnetic field configuration purely in the plane of the shock, with $\xi\rightarrow 0$, upper limits on the polarization of the radio afterglow of GW\,170817\,/\,GRB\,170817A indicate the magnetic field structure behind the shock has a significant component in the radial direction, along the shock normal, and is closer to being isotropic in 3D, rather than being almost purely in the shock plane \citep{GG18, Corsi2018, Stringer2020, Laskar2020ApJ, Teboul2021}. In B24 we adapt the values found in \citet{GG18} for the limits on the structure of the magnetic field behind the shock to the terms of our model. Here we present polarization curves using the lower limit with $\xi=0.75$ for the wind-like medium with $k=2$ (dark yellow dash-dotted curves in Fig. \ref{fig:THJModelsOnAxis}). %\pb{The previous sentence is not clear to me - what part is done in B24 and what part is done here?}\jg{see the comment just above \S\,\ref{subsec:OnAxisJets}} 
As shown in previous works, the degree of polarization reduces when more 3D isotropic magnetic field structures are introduced, while the typical times remain similar (B24, \citealt{Birenbaum2026}). Such models may very well reflect the nature of GRB afterglows for which optical polarization is measured.

\begin{figure}
	\includegraphics[width=\columnwidth]{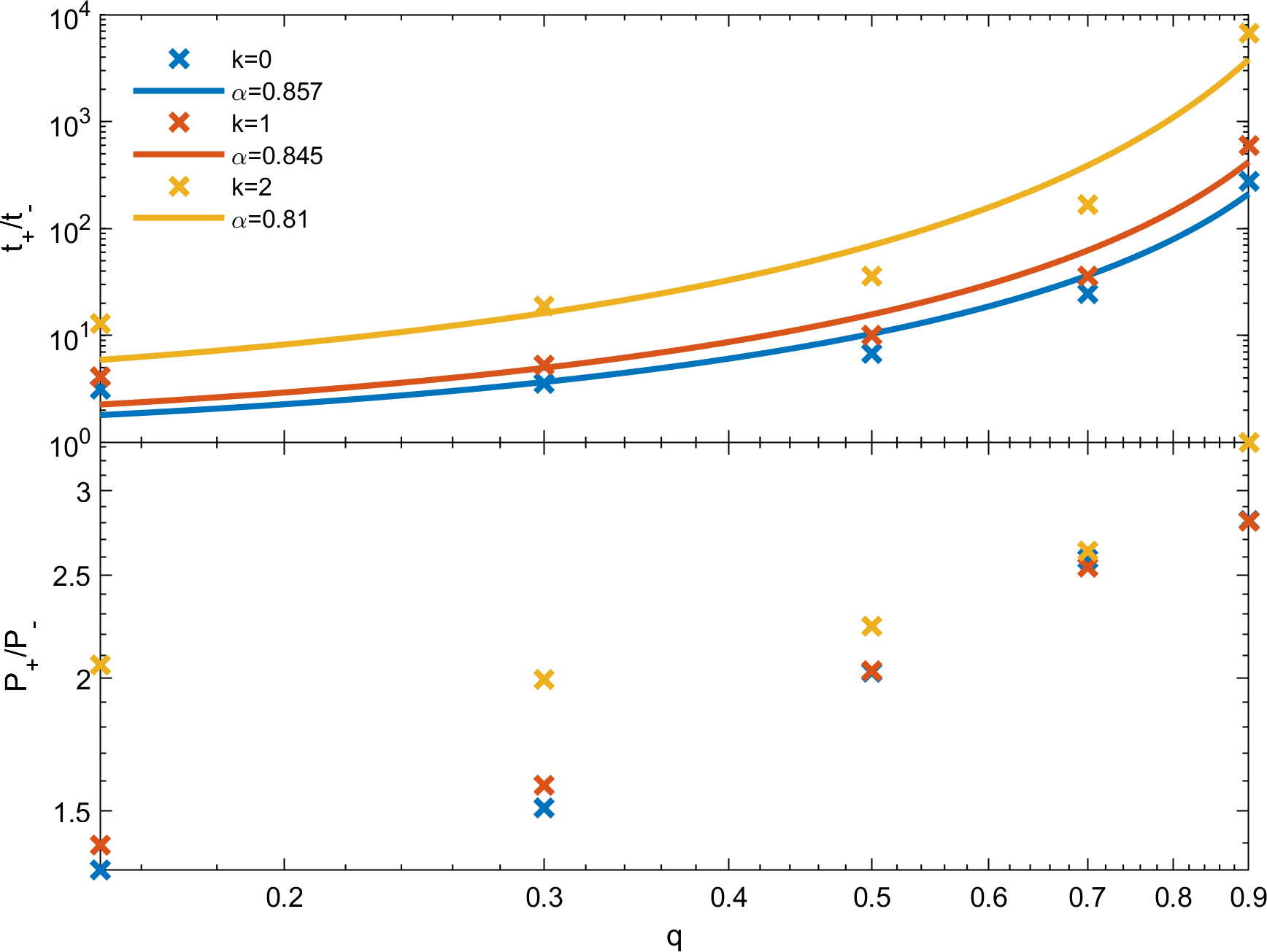}
    \caption{\textit{Upper panel: }Time ratio of the positively and negatively signed polarization peak times (marked $t_{+}$ and $t_-$ respectively) of on-axis top hat jet models, plotted as function of the normalized viewing angle $q=\frac{\theta\obs}{\theta_c}$%\pb{Remind the reader what $q$. Some readers will only look at abstract, figure captions, and conclusions, so those have to be relatively self-contained if possible.}
    , plotted in 'x' signs from the models considered in this work. The colors correspond to the power-law index of the external medium density profile. The solid lines, following the trends of the data points from the calculations made in section \ref{subsec:OnAxisJets}, are fitted according to the model developed in Eq. \ref{eq:PolTimeRatio}, with the corresponding value of $\alpha\left(k\right)$ given in the label for each value of $k$. \textit{Lower panel:} Ratio of the polarization peak heights for on-axis top hat jets as function of the normalized viewing angle $q$. The ratios of the wind-like model with $k=2$ are above the rest of the models for all values of $q$. The data points corresponding to $q=0.9$ with $k=2$ are plotted according to a slightly modified afterglow model in order to keep the first polarization peak in PLS G.}
    \label{fig:PolPeakTimeRatio}
\end{figure}

\begin{figure}
	\includegraphics[width=\columnwidth]{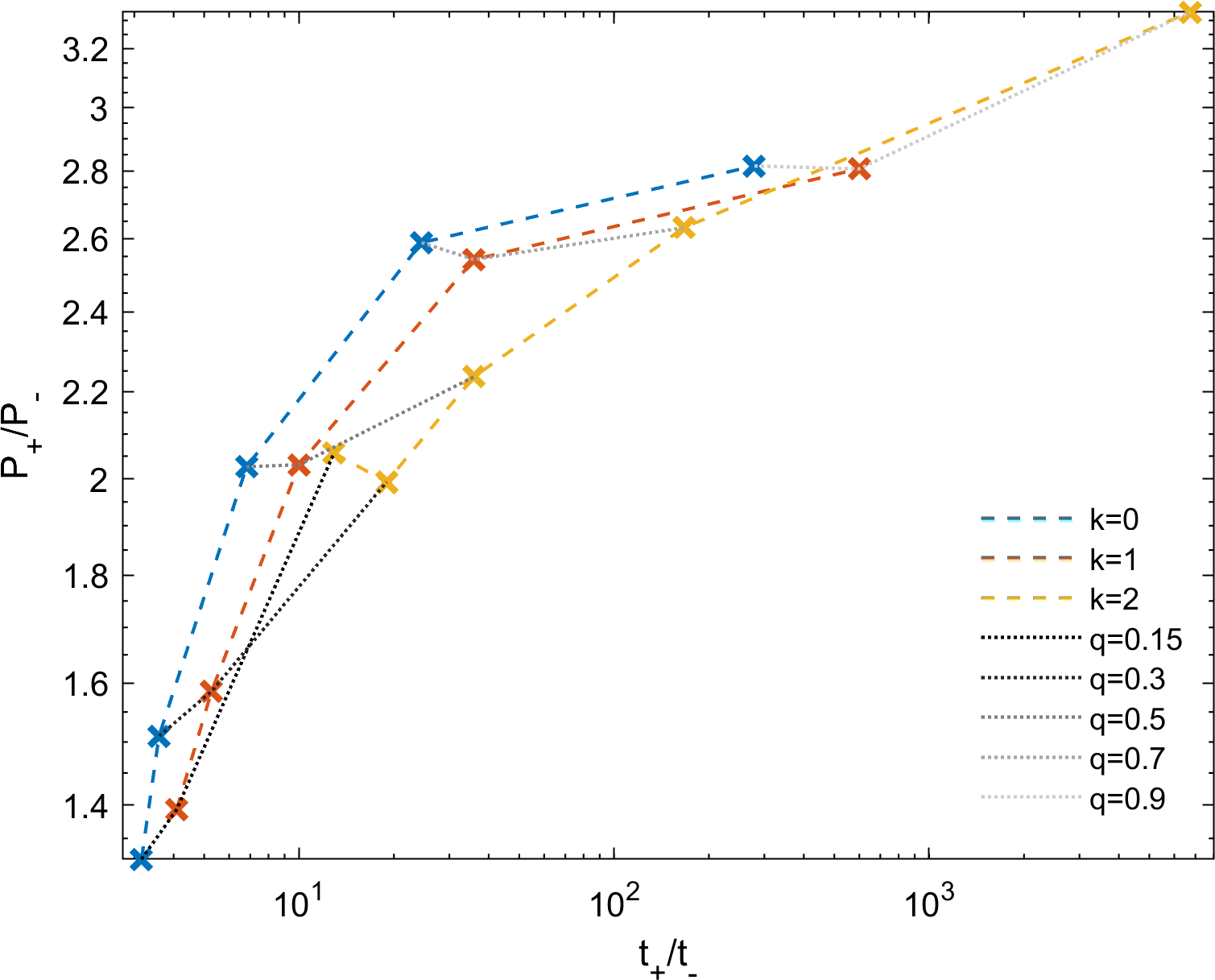}
    \caption{Two-dimensional parameter space for the polarization peak ratios presented in both panels of Fig. \ref{fig:PolPeakTimeRatio}. The points are connected by both equal-$k$ lines (dashed colored lines) and equal-$q$ lines (dotted gray lines). The data point corresponding to $q=0.9$ with $k=2$ are plotted according to a slightly modified afterglow model in order to keep the first polarization peak in PLS G. }
    \label{fig:2DSpace}
\end{figure}

 We mark the polarization peak times for an on-axis top hat jet model with $t_{\pm}$, where the minus (plus) sign corresponds to the first (second) polarization peak. In the Blandford-Mckee (BM; \citet{1976PhFl...19.1130B}) deceleration phase, $\Gamma\propto t\obs^{\frac{k-3}{2\left(4-k\right)}}$. 
 The angular radius of the emitting ring at these times is inversely related to the Lorentz factor $\tilde{\theta}_{\pm}\propto\Gamma_{\pm}^{-1}$, where it is measured from the LOS.
 The disappearance of the polarized ring beyond the jet edge causes the rise of polarization in on-axis top hat jets (\citealt{Birenbaum2021}; Figs. \ref{fig:THJq07WithMapsk0}-\ref{fig:THJq07WithMapsk2}). 
 However, the angular radius of the polarized ring becomes wider as the value of $k$ grows, causing  the effective viewing angle to become smaller, as the ring reaches the symmetry axis of the jet at earlier times. We parametrize this effect with the parameter $\alpha(k)$.
 During the first polarization peak, about a quarter of the polarized ring disappears beyond the jet edge and the angular scale of the ring is 
(e.g. \citealt{Birenbaum2021}),
\begin{equation}
    \tilde{\theta}_{-}=\theta_\text{c, eff}-\theta_{obs}=\theta_\text{c, eff}\left(1-\alpha(k)\cdot q\right),
\end{equation}
while during the 2nd polarization peak, when only a quarter of the polarized ring is visible, we can express:
\begin{equation}
    \tilde{\theta}_{+}=\theta_{\text{c, eff}}+\theta_{obs}=\theta_{\text{c, eff}}\left(1+\alpha(k)\cdot q\right).
\end{equation}
Using $\frac{t_+}{t_-}=\left(\frac{\Gamma_+}{\Gamma_-}\right)^{\frac{2\left(4-k\right)}{k-3}}$, we can get the analytical approximation for the time ratio of the polarization peaks:

\begin{equation}
\label{eq:PolTimeRatio}
    \frac{t_+}{t_-}=\left[\frac{1+\alpha(k)\cdot q}{1-\alpha(k)\cdot q}\right]^{\frac{2\left(4-k\right)}{3-k}},
\end{equation}
where the value of $\alpha(k)$ is of order unity and is found by fitting this expression to the time ratio of the polarization peak. 
These fits are shown in the upper panel of Fig. \ref{fig:PolPeakTimeRatio} in solid lines for the different media the afterglow forward shock propagates into. $\alpha(k)$ becomes smaller for larger $k$, encapsulating the fact that the polarized ring becomes wider at these values.
In the lower panel of Fig. \ref{fig:PolPeakTimeRatio}, we show the ratio between the heights of the two polarization peaks, where $P_-$ represents the negative polarization peak and $P_+$ is the positive one. As the jet symmetry axis becomes more misaligned with the line of sight, the positive polarization peak becomes higher compared to the negative one for almost all cases considered. %\sout{This ratio is also shown to be higher for $k=2$ for all viewing angles considered in this work.} 
%In addition, the gap between the different external density profiles becomes smaller as the value of $q$ grows and the various models converge to the maximal polarization the combination of the viewing angle and magnetic field structure are able to produce. 
In Fig. \ref{fig:2DSpace} we present two observables: the ratio of the polarization peak heights versus the polarization peak time ratios, in a 2 dimensional space, along with equal-$k$ lines (colored dashed lines) and equal-$q$ lines (dotted gray lines). While the range in $\frac{t_+}{t_-}$ spans three orders of magnitude, the area of the parameter space occupied by the models explored in this work is quite narrow. This manifests as higher peak polarization ratios having longer time scales. The implication is twofold: (i) The values of $k$ and $q$ can in principle be inferred from the observables $\frac{t_+}{t_-}$ and $\frac{P_+}{P_-}$. (ii) The model is falsifiable: there are regions in the observable 2D space that are inaccessible for any $k,q$ combination.

%\jg{maybe we can define the median $R_\perp$ or $\theta$ containing half of the contribution to the total flux and show how it increases with $k$, implying an increase in $\theta_{\rm c,eff}$ and a decrease in $q_{\rm eff}=\theta_{\rm obs}/\theta_{\rm c,eff}$ and $\alpha(k)=q_{\rm eff}/q$}

%\gb{Go over this and restructure and cut out text as fig. 9 is the important result}
While light curve models can constrain the viewing angle and the density profile of the external medium, they can be degenerate and sometimes several different models can fit the same light curve. We suggest using the time ratio between the two polarization peaks, as well as their height ratio, in order to uplift the degeneracy. While the peak polarization is highly dependent on the shock-generated magnetic field structure and can be indicative of it, these ratios are independent of it, as the shape of the polarization curve is conserved under changes to the $\xi$ parameter. %\sout{ Managing to observationally capture both opposing polarization peaks  with a well sampled light curve, that allows constraining the slope of the external medium density profile via light curve slopes, can assist in constraining the viewing angle to the system using the expression in Eq. \ref{eq:PolTimeRatio} and parameter space presented in Fig. \ref{fig:PolPeakTimeRatio}.} %\jg{If two $90^\circ$ rotated peaks are at all observed...} 
Constraining the viewing angle to the system often relies on the identification of the jet break and timing of it, and wind-like models have been invoked in the past for light curves with very gradual jet breaks that are hard to place on the light curve (e.g., \citealt{Gill2023}). %\sout{However, a good handle on the polarization peak times, \gb{their magnitudes} and $k$ from observations can make constraining the viewing angle to the system simpler.}
Managing to observationally capture both polarization peaks and placing their respective ratios in the 2D representation of the observable parameter space (Fig. \ref{fig:2DSpace}) can constrain the normalized viewing angle and identify the external medium density profile the shock propagates into simultaneously, in a way that is independent of the chosen afterglow model.

\begin{comment}

 We can thus express the ratio between these critical times as
 \begin{equation}
     \frac{t_+}{t_-}=\left(\frac{\Gamma_+}{\Gamma_-}\right)^{\frac{2\left(4-k\right)}{k-3}}=\left(\frac{1}{1-q}\right)^{\frac{2\left(4-k\right)}{3-k}}
 \end{equation}
\end{comment}
\subsection{Jets viewed off-axis ($q\ge 1$)}
\label{subsection:OffAxisJets}
We explore next the polarization signature from off-axis jets propagating into power-law external density profiles.
%and analyze how propagation into \sout{non-trivial} external density profiles affects the observed polarization signature.
We expand on the results for top hat jets, shown in section \ref{subsec:OnAxisJets}, to the regime of jets observed off-axis in section \ref{subsection:OffAxisTHJ}. We then proceed to analyze the polarization signatures of off-axis power-law structured jets that are propagating into power-law external density profiles in section \ref{subsection:OffAxisPLJ}.
\subsubsection{Top hat jets}
\label{subsection:OffAxisTHJ}
While top hat jets may not be the most realistic description of relativistic jet afterglows, understanding these systems and how their structure produces their light and polarization curves helps build intuition and acts as a comparison point for more realistic structured jets (section \ref{subsection:OffAxisPLJ}). 

In Fig. \ref{fig:THJModelsOffAxis}, we model the polarization and light curves of a top hat jet, viewed with off-axis ($q=2$), with varying values of $k$. The light curve evolution of an off-axis jet is explained in previous works (e.g. \citealt{Granot2002, Lamb2017, Beniamini2020,Teboul2021}). %\textcolor{red}{When a top hat relativistic jet is observed at a viewing angle beyond its core opening angle ($\theta\obs>\theta_{\text{c}}$), initially emission from the jet core is beamed away from the observer. As the core is highly relativistic,  they will observe a very low initial flux, as there is no material moving along the LOS. Since the core of the jet is decelerating, its emission becomes less beamed over time, leading to a rise in observed flux.} 
Since the narrow core of the jet is misaligned with the LOS, it is observed as a point source with an ordered magnetic field, due to synchrotron radiation being best observed from the components of the magnetic field lines that are perpendicular to our LOS. 
Therefore, the observed emission from the core is highly polarized, and the rise in flux will be accompanied by a rise in polarization. Eventually, the relativistic jet core decelerates enough that it becomes fully visible to the observer, culminating in a peak of the light and polarization curves. %\textcolor{red}{Beyond this point, we are not seeing new parts of the jet core and it continues to further decelerate, resulting in the typical power-law decline of the light curve \citep{GG18,Beniamini2020,Beniamini2022}.} %\jg{I think that the main effect governing $\Pi(t_{\rm obs})$ for $q\gtrsim2$ is the evolution in $\theta'$ as $\Gamma(t_{\rm obs})$ decreases, since for $q\gtrsim2$ the point source approximation works well, and the local $\Pi[\theta'(t_{\rm obs})]$ governs the observed $\Pi(t_{\rm obs})$} 
The break in the light curve is temporally close to the peak of the corresponding polarization curve (see Appendix~\ref{app:TimeRatios} for more details about this proximity).
%\footnote{details in appendix \ref{app:TimeRatios}.}.
From this point on, the observer is able to also see emission from all components of the magnetic field behind the shock, leading to an eventual reduction of polarization due to increased cancellations following the peak. %\jg{at $t>t_j$ the whole jet is visible, but all components of the local magnetic field exist in the visible region at all times (also at $t<t_j$); I think that the reason for the reduction in $\Pi(t_{\rm obs})$ at $t>t_j$ is the decrease in $\theta'(t_{\rm obs})$ and correspondingly in the local emitted $\Pi[\theta'(t_{\rm obs})]$.}
The polarization will have a fixed sign throughout the temporal evolution of the system, either along (for $\log\xi<0$) or perpendicular to (for $\log\xi>0$)
the favored direction in the system that connects the LOS with the jet symmetry axis (\citealt{Granot2003a,Rossi2004,Birenbaum2021,Birenbaum2026}; B24). %\pb{The discussion above mixes somewhat the reasoning behind the shape of the light-curve and that of the normalized polarization curve.} \gb{should I explain them separately?}\gb{less elaboration on off axis LC evolution as it has been covered in lost of previous papers}

This behavior is evident in Fig. \ref{fig:THJModelsOffAxis}. %\sout{, we can see that indeed all polarization signatures for off-axis top hat jets in different external density profiles behave in a fashion similar to that described above}. 
However, there are several differences between the examples shown. While the light curves of the $k=0,1$ (blue and red models respectively) models show a peak in the light curve, the $k=2$ (yellow model) case only shows a declining flux. This is due to the core reaching its deceleration radius at earlier times due to its early interaction with the high density medium, which acts to decelerate it quickly. Therefore, the early rising phase of the observed flux occurs before the times shown in Fig. \ref{fig:THJModelsOnAxis}. The early rise in observed polarization of the yellow curve supports that claim as well, as it indicates light from the jet core is already observed at this point (see also \citet{Lan2023}). The shape of the polarization peak depends on $k$. This can be explained by considering the changes in the width of the polarized ring around the LOS, as detailed for $q<1$ in section \ref{subsec:OnAxisJets}. Since the jet structure is a uniform top hat jet, when light from the core is observed, it is in the form of a ring around the LOS, the center of which now lies outside of the jet core. For $k=0,1$ there is a double peaked polarization signature, where both have the same sign (see also the $k=0$ case presented in \citealt{Rossi2004}). The ring around the LOS is thin and the first polarization peak occurs when the edge of the polarized ring reaches the closer end of the core of the jet \footnote{see panel (a) of Fig. \ref{fig:THJq07WithMapsk0} for an example for how the ring interacts with the jet edge}. %\jg{A figure showing the difference between $k=0,\,1,\,2$ described here will really help} 
Thus, most of the observed light arrives from a very narrow region across which the polarization vector is ordered, leading to the emergence of the first highly polarized peak. Following this point, the jet core continues to decelerate %, which allows us to see more parts of it as relativistic beaming weakens. This manifests as more parts of the polarized ring becoming visible, allowing for some cancellations to arise, which lowers the observed polarization slightly. 
until it reaches the geometrical light curve break time after which the polarization and light curves will exhibit post-jet-break temporal evolution, as explored in section \ref{subsec:OnAxisJets}.
The height of the polarization peaks is a direct result of the ring width. Since the ring is wider for the $k=1$ case (red lines), its second polarization peak is higher than the corresponding one for the $k=0$ case (blue lines) as the part over which the polarization vector is ordered is wider and dominates the emission region (as discussed in section \ref{subsec:OnAxisJets} for $q<1$). 
For the wind-like external medium ($k=2$; yellow lines), while the dynamical time between the two peaks is longer, they are also more smeared %\pb{This is misleading - the time between the centers of the two peaks is actually longer, they are just much more smeared} 
due to the increased width of the ring, causing them to merge into one peak. Since it is the same lobe of the polarized ring that dominates the emission though the evolution of the system, the polarization remains at the same sign throughout for all cases explored in this part. 

We also present a wind-like external medium case ($k=2$) with a more realistic magnetic field structure behind the shock, with the observationally motivated value $\xi=0.75$, which naturally exhibits a similar polarization signature with reduced magnitudes. %\pb{This paragraph is essentially contained in the figure caption. Either remove completely from this instance or add some clear bottom line.}

\subsubsection{Structured jets}
\label{subsection:OffAxisPLJ}
\begin{figure*}
  \centering
  %\subfigure%(a)
    {\includegraphics[width=\columnwidth]{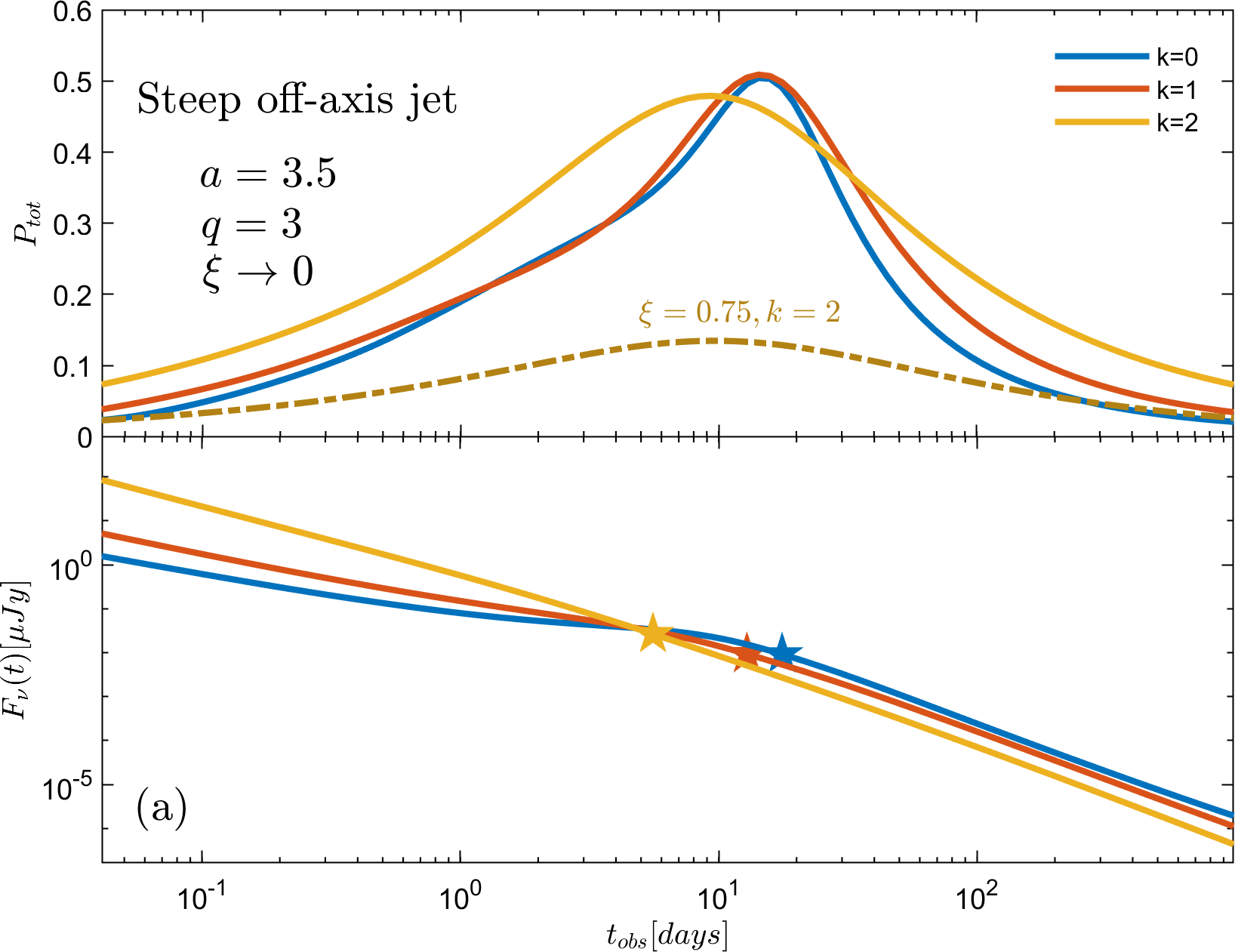}} 
 % \subfigure%(b)
  {\includegraphics[width=\columnwidth]{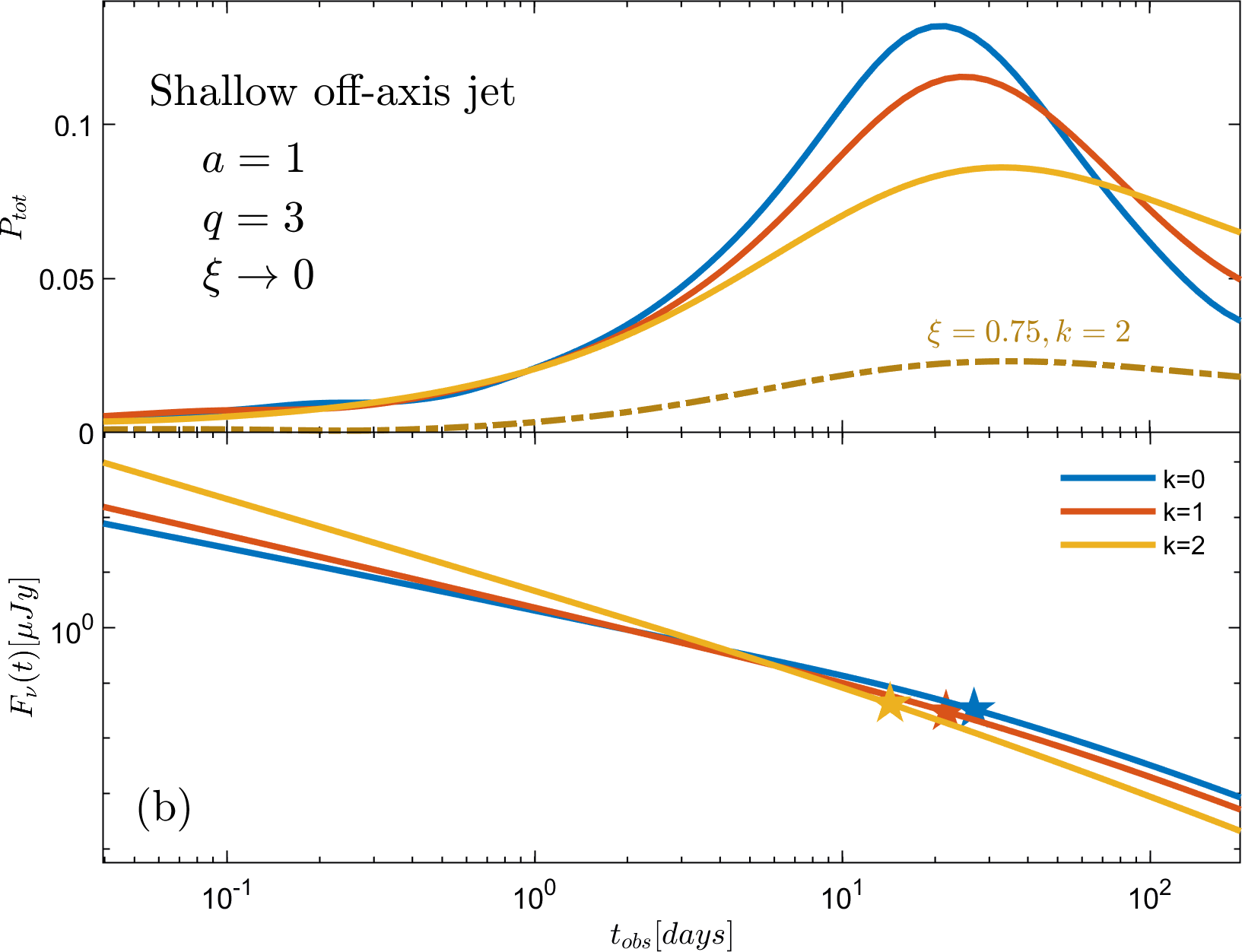}} 
  \caption{Observed polarization degree (\textit{upper panels}) and flux (\textit{lower panels}) at observer frequency $\nu=10^{15} \text{ [Hz]}$ (PLS G) as function of time for structured jet models, with a random magnetic field structure, confined to the shock plane ($\xi\rightarrow 0$) and an off-axis viewing angle of $q=3$. The various values of k represent different power-law indices of the external medium density profile. The light curve break is marked
with a star in the lower panels. \textit{Panel (a)} presents the observables from a shallow jet with $a=1$. \textit{Panel (b)} presents the same observables from a steep jet with $a=3.5$. The dash-dotted dark yellow lines present the polarization curves for cases considering a more realistic magnetic field
configuration behind the shock with $\xi=0.75$. }
 
\label{fig:P2Together}
\end{figure*}
%\sout{While the analysis of afterglows from impulsive relativistic jets with top hat jet structure presented above is useful for gaining intuition for the temporal evolution of their polarization and light curves,} 
Observational \citep{Mooley2018, GG18, Gill2020, Gill2023, O'Connor2023} and theoretical \citep{Granot2002, Nakar2020,Gottlieb2021} evidence indicate that GRB jets posses extended angular structure beyond the narrow jet core. The effects of an extended jet structure on the polarization properties were explored in depth in previous work (\citealt{Rossi2004, GG18,Gill2020, Teboul2021, Birenbaum2026}; B24). These studies have shown that shallow jet structures\footnote{where the energy in the jet extended structure is comparable to that in the jet core ($a\lesssim 2$, see \citealt{Beniamini2022} for more details)}, which are inferred from the multi-wavelength afterglow light curves of GRB 221009A \citep{O'Connor2023, Gill2023}, also agree with the optical and X-ray limits in the polarized regime (B24). Modeling the polarization of slightly off-axis structured jets  
along with their light curves, also has potential for differentiating between intrinsic origin scenarios for orphan afterglow candidates \citep{Birenbaum2026}. Such structured, off-axis jets can also describe delayed radio emission from jetted TDEs \citep{Cendes2026}. Detection of radio polarization from these systems can potentially differentiate them from delayed outflows, also suggested to power this emission (e.g., \citealt{Alexander2026}.)%\pb{Here you can also mention delayed outflows proposed in radio TDEs.}

In this section we present the effects of such afterglow forward shocks propagating into a stratified medium. 
In Fig. \ref{fig:P2Together}, panel (a), we present an off-axis ($q=3$) jet with a steep structure ($a=3.5$) and varying power-law indices of the external density profile. As shown in \citet{Birenbaum2026}, the polarization signature of these off-axis steep jets consists of a single polarization peak with the polarization sign remaining the same throughout the evolution of the system in a similar fashion to the top hat jet case presented in section \ref{subsection:OffAxisTHJ} above. In addition, the light curve shows prominent breaks which are due to the jet core being the main contributor to the observed flux after the jet break. Since most of the jet energy is in the jet core, the system becomes highly asymmetrical during the polarization peak, leading to polarization levels that approach the maximal polarization level allowed for synchrotron radiation \citep{Rybicki1979}. %\jg{for a large $q$ the observed image becomes increasingly asymmetric during the rise to the lightcurve peak \citep[see Fig. 3 of][]{Beniamini2022} while at $t\gtrsim t_{\rm peak}$ it remains highly asymmetric, dominated by the jet core that can be approximated as a point source, and the polarization decreases due to the evolution of $\theta'$ and $\Pi[\theta'(t_{\rm obs})]$} 
In this case, the wind-like medium ($k=2$; yellow lines) has a slightly lower peak polarization level compared to the uniform medium case ($k=0$; blue lines). %initially the observed emission is dominated by light from material moving along the LOS that originates from the extended structure of the jet. A rise in polarization occurs when light from the jet core starts to dominate the observed emission, which culminates in a peak when the core is fully visible. \pb{This also repeats the explanation given previously in the draft}

In dash-dotted dark yellow lines we present the most realistic configuration with both wind-like external medium density profile ($k=2$) and a more 3D isotropic magnetic field structure ($\xi=0.75$). While the shape of this polarization curve is similar to that of the $\xi\rightarrow 0$ case, its magnitude is reduced greatly due to the changed structure of the magnetic field.

In Fig. \ref{fig:P2Together}, panel (b) we repeat the analysis performed above for shallow jets. We select the same off-axis viewing angle with $q=3$, vary the jet structure to a shallow jet with $a=1$ and explore the changes in the observed flux and polarization when the power-law index of the external medium density profile changes. These systems also show a single polarization peak with modest levels and constant sign throughout the evolution (B24). Following the time of the polarization peak, the energetic extended structure of the jet starts contributing more to the observed emission, which acts to increase the symmetry of the system and lower its observed polarization. 
This effect also makes the break in the light curve (marked with stars) very gradual.
The magnitude of the peak polarization level for the wind-like medium is reduced compared to the uniform density external medium, similar to some of the cases previously explored in this work. While the peak polarization for a steep jet has a very weak dependence on $k$, being very similar (varying by only a factor of $\sim$1.06) for $0\leq k\leq2$, the shallow jet configuration shows a range in peak polarization levels that spans a factor of $\sim 1.6$.%\jg{The problem is that for any afterglow we can measure one peak time, and it has a larger dependence on other unknown parameters, e.g. $q$, $E_c$, or density normalization.}

We also include a version of this model with a more realistic structure for the magnetic field behind the shock for the wind-like external medium ($\xi=0.75$; dash-dotted dark yellow line). As discussed above, considering a more 3D isotropic magnetic field structure behind the shock lowers the overall magnitude of the polarization signature.

While steep jets hold less overall energy in our setup\footnote{as $E_c$ is the same for all setups explored in this work.}, due to their structure and the radius at which the external density profiles explored in this work cross, they seem to provide the best chances for measured polarization as their polarization curves peak at earlier times, where the observed flux is also higher. Steep off-axis jets would also produce higher polarization levels close to the light curve break time, even in more realistic cases with wind-like media and a more 3D uniform magnetic field structure ($\xi=0.75$; dark yellow lines). Based on the results of this section, we conclude that for realistic off-axis jets that take into account the possibility of a structured jet and non-constant external medium, polarization would be most easily observed for steep jets.

\section{Discussion \& Conclusions}
\label{sec:Discussion}
In this paper we model the observed polarization signature and flux from impulsive jet afterglows that propagate into external media with power-law density profiles behaving as $\rho\propto r^{-k}$, with $0\leq k\leq 2$. Such models are applicable to long GRBs, which are the majority of GRBs for which afterglow polarization has been measured, and jetted TDEs. 

While the afterglow model contains many parameters, the shape of the polarization curve will depend mainly on the shape of the external density profile, the geometrical setup of the system (i.e. viewing angle and jet structure), the part of the spectrum the observed frequency probes\footnote{PLS G in this work.} and the degree of anisotropy of the shock-generated magnetic field. Previous works have shown that the polarization signature and features change in a smooth manner with the latter parameters, where higher values of $q$ and $a$ produce higher peak polarization levels, while varying the stretching parameter $\xi$ can change the sign of the polarization signature (i.e. if the local polarization is in the radial or azimuthal direction w.r.t the line of sight), where the sign change occurs at $\xi=1$ \citep{Granot2003a,Rossi2004,Birenbaum2021, Birenbaum2024, Birenbaum2026}. Increasing the power-law index of the external density profile $k$, which changes it from uniform to declining with radius from the origin, causes a reduction in observed polarization for most cases studied in this work, where wind-like models will produce lower observed polarization signatures compared to models involving a uniform medium. Such change will also cause a slower
temporal evolution of the polarization curve, prolonging the time scale between its features. In addition, an increased value of $k$ can also make the geometrical light curve break (i.e. jet break for an on-axis observer) more gradual and difficult to determine from observations.

Identifying the time of the geometrical light curve break %(i.e. jet break for on-axis jets) 
is currently one of the main methods to determine the opening angle of the jet and the viewing angle to the system. Therefore, it is essential to develop additional methods to constrain these quantities for cases where it is not possible to do so based on light curve analysis alone. For on-axis top hat jets, managing to measure the two oppositely signed (i.e. with a polarization angle differing by 90$^\circ$) polarization peaks in terms of their magnitudes and times can simultaneously constrain both the normalized viewing angle to the system $q$ and the value of $k$. Since the parameter space occupied by the models is quite narrow, this method can be used as a self-consistency test of this model. This approach is also independent of the structure of the shock-generated magnetic field (i.e. of the value of $\xi$).

%5. Application 2: This is applicable for GRBs for which polarization is measured (notably, the BOAT). mention again that most grbs with polarization measurements are long grbs and that we must include the possibility of wind

The lack of an easily identified geometrical light curve break for the afterglow of GRB 221009A, in addition to its extreme $E_{k,\text{iso}}$ \citep{Burns2023}, prompted afterglow models that invoke a shallow jet structure for this system \citep{Gill2023,O'Connor2023}. In addition to the jet structure, one of the models also explained the lack of a noticeable jet break with the afterglow shock propagating into a wind-like medium with $k=2$ \citep{Gill2023}. While this burst was bright enough to place upper limits on its X-ray polarization at $3.5$ days post-burst \citep{Negro2023}, in B24 we show that measuring the polarization at earlier times could have constrained the stratification of the external medium of this burst, by differentiating between the \citet{Gill2023} and the \citet{O'Connor2023} models.
This highlights the importance of allowing for a stratified external medium when modeling long GRB afterglow observations.

%4. Application 1: This is applicable for TDEs

Afterglows from steep jets have also been shown to describe delayed radio emission from possibly jetted TDEs \citep{Cendes2026}. Combined with the stratified media some of these events occur in \citep{Sfaradi2024,Granot2026}, polarization models of impulsive jet afterglows propagating into power-law external density profiles can help in differentiating between different scenarios for the origin of delayed radio emission in TDEs (i.e. off-axis relativistic jets and delayed outflows; for more details, see \citealt{Alexander2026}).

%6. sum it up: realistic models that reflect the reality of GRBs for which polarization is measured and possible jetted TDEs.

Our findings stress the importance of modeling and interpreting the polarization and light curves jointly. In this work we add the possibility of a stratified external medium as a factor that relates between the models and the physical nature of both TDE and GRB afterglows. Models involving propagation into stratified media can explain the lack of a clearly observed jet break for GRB afterglows, as well as differentiate between different origins for delayed radio emission from TDEs.

\begin{acknowledgements}
PB and GB are supported by a grant (no. 1649/23) from the Israel Science Foundation.
PB is also supported by a grant (no. 2020747) from the United States-Israel Binational Science Foundation (BSF), Jerusalem, Israel and by a grant (no. 80NSSC 24K0770) from the NASA astrophysics theory program.
\end{acknowledgements}
\bibliographystyle{aa}
%\bibliography{aanda}
\bibliography{library}
 \begin{appendix}
\section{Proximity of critical times}
\label{app:TimeRatios}

\begin{figure}
	\centering
    \includegraphics[width=\columnwidth]{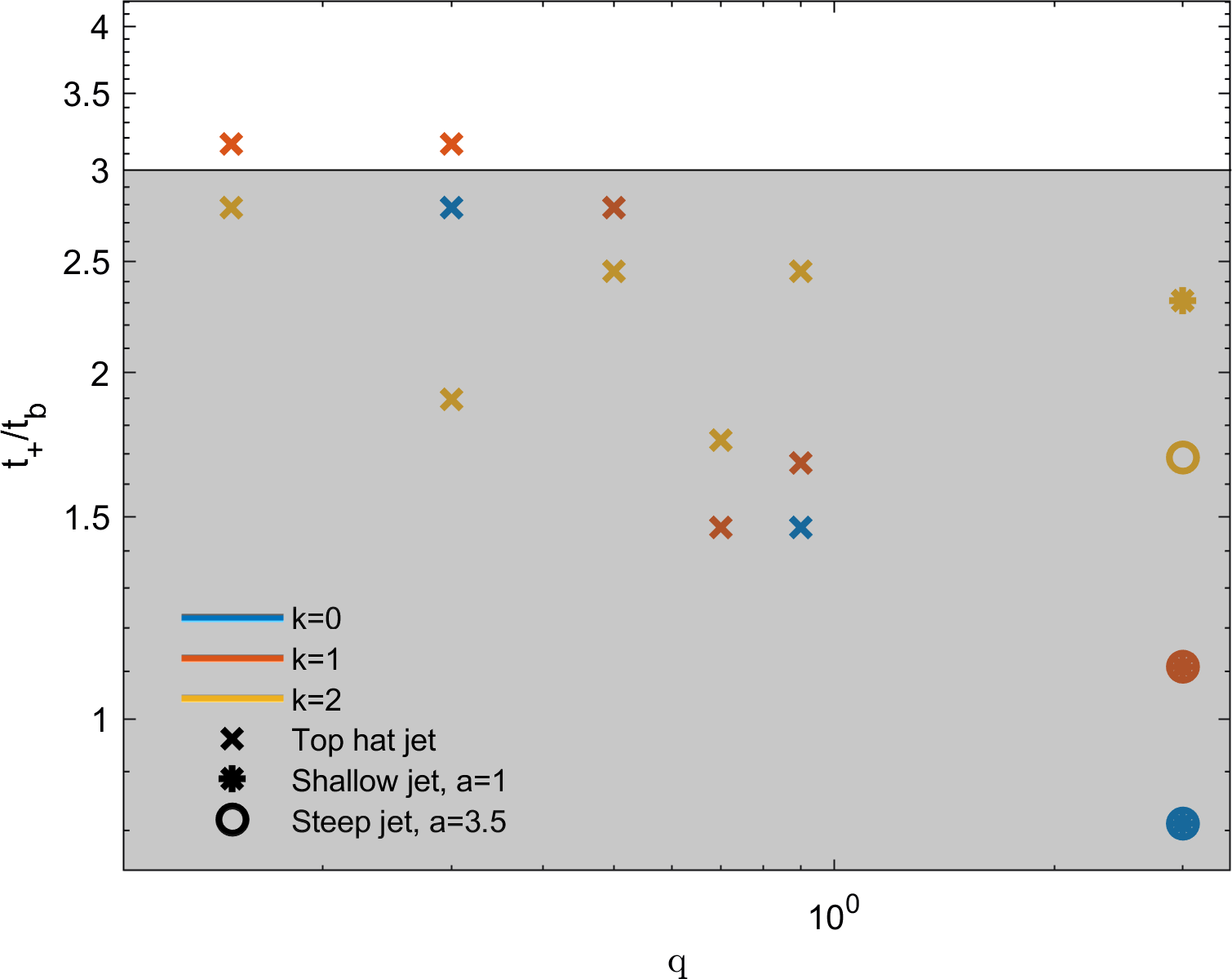}
    \caption{Ratio between the main polarization peak times $t_{+}$ and light curve break times $t_b$ as function of the normalized viewing angle $q$ for all setups considered in this work. We can see that for most models, these two times are within factor 3 (marked by the gray shaded region) of one another or close to that. }
\label{fig:TimeRatioPlot}
\end{figure}
Following the analysis presented above, we proceed to explore the temporal relationship between the light curve break and the polarization peak. In previous works by B24 and \citet{Birenbaum2026}, it was shown that these two times are geometrically related and demonstrate a temporal proximity of up to a factor of $3$ between them. Such proximity cements the idea that polarization measurements conducted close to the break time of the light curve probe the maximal polarization level the given geometrical setup of the afterglow can produce, and should be prioritized. In Fig. \ref{fig:TimeRatioPlot} we present the analogous quantification for the cases considered in this work and plot the time ratio of the polarization peak to the time of the light curve break as function of the viewing angle $q$. For the on-axis jet cases, we consider the 2nd, positive, polarization peak. The light curve break time is determined by fitting power-law functions to the asymptotic parts of the light curve and finding their crossing time. 

We can see that for almost all setups considered in this work, these two critical times are within factor 3 of one another, shown as a shaded gray region across the parameter space. We choose to exclude the off-axis top hat jet model with varying values of $k$, that feature a double peaked polarization signature which remains at the same sign throughout. This makes the relation between the light curve break time and the polarization peak less straight-forward compared to definitive cases with either oppositely signed polarization peaks or a clear single-peaked polarization curve.  The cause for this unique signature is discussed at length in section \ref{subsection:OffAxisTHJ}.

In most of the cases, the light curve break precedes the polarization peak, where the only setups with this time ratio being below unity are the off-axis structured jets that propagate into a uniform external medium profile ($k=0$; blue overlapping '*' and 'o' signs). This suggests that the inclusion of a non-trivial external medium delays the evolution of the polarization more than it does the corresponding light curve. This can be explained by the widening of the polarized ring for values of $k>0$, which smear the evolution of the polarization signature to longer time scales while the evolution of the light curve is more affected by the average radius of the ring that contributes more to the observed flux. Overall, the critical times of the models considered in this work are mostly within factor of three of one another, consistent with the results shown in previous works (B24 and \citet{Birenbaum2026}), further cementing the relationship between the main polarization peak and the geometrical light curve break.

\section{Analytical expressions for the EATS in power-law external media}
\label{app:AnalyticalEATS}

\begin{figure}
	\centering
    \includegraphics[width=\columnwidth]{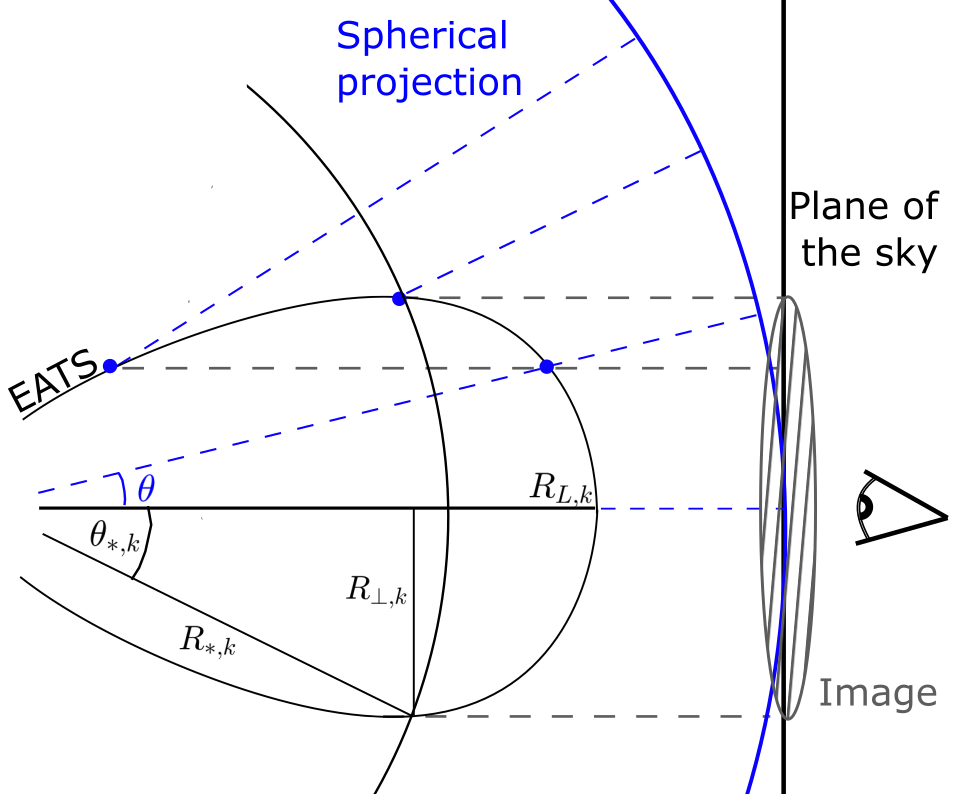}
    \caption{\textit{Black lines:} shape of the EATS from an on-axis forward shock at $R_{L,k}$, similar to that shown in Fig. \ref{fig:EATSAllk}, with critical radii and angle marked. We plot this for a perfectly on-axis jet, with $q=0$, where the LOS and the jet symmetry axis are aligned. \textit{Blue lines}: spherical projection of the emission and polarization from the forward shock used in our formalism to plot the polarization maps (see Figs. \ref{fig:THJq07WithMapsk0}-\ref{fig:THJq07WithMapsk2}). The observed flux and polarization from each polar angle $\theta$ from the symmetry axis is mapped onto this projection. \textit{Gray lines}: observed image, plotted from the EATS, where the edge of the image corresponds to an angular radius of $\theta_{*,k}$, where other angles on the EATS contribute twice. While this formalism is not used in this work, it is included in this plot for completeness.}
\label{fig:ProjectionSetup}
\end{figure}

We start by expanding the analytical expressions for the equal arrival time surface (EATS) given for the combination of radii and angles on the shock that contribute to the afterglow emission at a specific observer time $t\obs$, generalizing the expressions given in \citet{Sari1998Egg} and \citet{Granot1999ApJ} for a general external medium. While the formalism used to evaluate the afterglow dynamics and emission in this work already includes a version of this generalization, which is naturally obtained by solving the observer time evolution of the Lorentz factor $\Gamma$ and shock radius $R$ (\citealt{GG18}, B24), we also present the analytical generalization from first principles. For a decelerating relativistic fireball, the Blandford-Mckee (BM; \citet{1976PhFl...19.1130B}) deceleration profile scales as $\Gamma\propto R^{\frac{k-3}{2}}$ for adiabatic evolution. This can provide the time interval an observer measures between two photons on the line of sight (LOS), where the first is measured when the shell is launched, and the second when it is at a radius $R$ \citep{Nakar2007}:
\begin{equation}
    t\obs=\frac{R}{2\left(4-k\right)c\Gamma^2}
\end{equation}
We can now relate the lab time that contributed to the emission at a certain radius on the LOS through $t\obs=t_{\text{lab}}-\frac{r\mu}{c}$, where $\mu=\cos\theta$ and on the LOS $\mu=1$:
\begin{equation}
    R=\frac{ct_{\text{lab}}}{1+\frac{1}{2(4-k)\Gamma^2}}
\end{equation}
and at a general angle $\mu$ relative to the LOS we obtain:
\begin{equation}
    R\left(t\obs,\mu\right)=\frac{ct\obs}{1-\mu+\frac{1}{2\Gamma^2(4-k)}}
    \label{eq:R(t_obs)}
\end{equation}
When we mark the radius and Lorentz factor on the LOS with $R_{L,k}$ and $\Gamma_{L,k}$ respectively, we can express $\Gamma=\Gamma_{L,k}\left(\frac{R}{R_{L,k}}\right)^{\frac{k-3}{2}}$ and plug into Eq. \ref{eq:R(t_obs)}:
\begin{equation}
    1-\mu=\frac{1}{2\left(4-k\right)\Gamma_{L,k}^2}\left[\left(\frac{R}{R_{L,k}}\right)^{-1}-\left(\frac{R}{R_{L,k}}\right)^{3-k}\right]
\end{equation}
The vertical distance from the LOS can be expressed in terms of $R_{\perp,k}$ (see Fig. \ref{fig:ProjectionSetup}):
\begin{equation}
    R_{\perp,k}=R\sqrt{1-\mu^2}\approx \frac{R_{L,k}}{\sqrt{4-k}\Gamma_{L,k}}\sqrt{\left(\frac{R}{R_{L,k}}\right)-\left(\frac{R}{R_{L,k}}\right)^{5-k}}
\end{equation}
which peaks at $R_{*,k}/R_{L,k}=(5-k)^{-1/(4-k)}$ or $\frac{R_{*,k}}{R_{L,k}}=5^{-1/4}\approx0.669$ for $k=0$, at $4^{-1/3}\approx0.630$ for k=1 and $3^{-1/2}\approx0.577$ for $k=2$. %\jgout{leading to an increase in the maximal polar angle that contributes to the emission $\theta_*$ with the value of $k$ which obeys, $\theta_{*,2}>\theta_{*,1}>\theta_{*,0}$ (see Fig. \ref{fig:EATSAllk})} 
Angles corresponding to $\theta>\theta_*$ also contribute to the emission, and the ordering of $\theta_{*,k}$ for $k=0,\,1,\,2$ depends on the choice of $t_{\rm obs}$ and the density normalization.  %\jg{this is the known result $y_*=R_*/R_l=(5-k)^{-1/(4-k)}$} 
These results are consistent with those presented at \citet{GS02} and \citet{Granot2008}. Fig. \ref{fig:EATSAllk} shows realizations of the EATS, corresponding to the external density power-law profiles explored in this work in solid lines, evaluated at the same observer time. In Fig. \ref{fig:ProjectionSetup} we show the how flux from the EATS can be projected in different ways. One way, which is not used in this work, is by looking at the observed image on the plane of the sky, where at the edge of it, at $\theta_{*,k}$, the image experiences limb brightening \citep{Granot1999ApJ,Granot+2005,Granot2008} and the rest of the points on the EATS are mapped twice onto the plane of the sky (see gray lines Fig. \ref{fig:ProjectionSetup}). In panel (a) of Figs. \ref{fig:THJq07WithMapsk0}-\ref{fig:THJq07WithMapsk2} we demonstrate the results of a spherical projection of the EATS, where the flux contribution and local direction of the polarization angle at each angular cell on the shock are mapped once onto a spherical surface (see blue lines in Fig. \ref{fig:ProjectionSetup}). Such setup allows us to explore the interplay between the polarized bright ring and the edge of the jet that produces the well-known polarization signature of on-axis top hat jets \citep{Sari1999Pol,Nava2015a,Birenbaum2021}.

 \end{appendix}

\begin{comment} 

\end{comment}
\end{document}